\documentclass{article}
\usepackage{amsmath}
\def\R{{\bf R}}
\numberwithin{equation}{section}

\begin{document}

\title{Dynamical properties of models for the Calvin cycle}

\author{Alan D. Rendall\\
Max Planck Institute for Gravitational Physics\\
Albert Einstein Institute\\
Am M\"uhlenberg 1\\
14476 Potsdam, Germany\\
\\
Juan J. L. Vel\'azquez\\
Institute for Applied Mathematics\\
University of Bonn\\
Endenicher Allee 60\\
53115 Bonn, Germany}

\date{}

\maketitle

\begin{abstract}
Modelling the Calvin cycle of photosynthesis leads to various systems of
ordinary differential equations and reaction-diffusion equations. They
differ by the choice of chemical substances included in the model, the choices 
of stoichiometric coefficients and chemical kinetics and whether or not 
diffusion is taken into account. This paper studies the long-time behaviour of 
solutions of several of these systems, concentrating on the ODE case. In some 
examples it is shown that there exist two positive stationary solutions. In 
several cases it is shown that there exist solutions where the concentrations 
of all substrates tend to zero at late times and others (runaway solutions) 
where the concentrations of all substrates increase without limit. In another
case, where the concentration of ATP is explicitly included, runaway solutions 
are ruled out.
\end{abstract}

\section{Introduction}\label{intro}

Photosynthesis is a process which is of great importance for many reasons. It
is the ultimate source of the food we eat, the oxygen we breath and 
many fuels (fossil fuels and biofuels). For this reason it is clear that it
would be valuable to have a better theoretical understanding of this 
process and one way of approaching this task is to use mathematical models. 
The aim of this paper is to analyse dynamical properties of some of these 
models.

Photosynthesis can be split into two main parts. In the first part, called
the light reactions, energy is captured from light and the small molecules ATP 
and NADPH are produced. These provide sources of energy and reducing power, 
respectively, for the second part, the dark reactions. The name of the latter 
comes from the fact that they can take place in the dark. Molecular
oxygen is produced during the first part. In the second part carbon
dioxide is used to make carbohydrates. For this reason this part is also
known as carbon fixation. This paper is exclusively concerned with models for
the second part of photosynthesis. 

The models which are relevant describe reactions between different chemical
substances and also, in some cases, diffusion of these chemicals. The
resulting mathematical model is a system of ordinary differential equations
(ODE) if diffusion is not included and a system of reaction-diffusion 
equations if it is. The models studied in what follows are either taken 
from the papers \cite{grimbs11} or \cite{zhu09} or are closely related to the 
models in those papers. In all these cases the network of reactions modelled
contains a cycle and due to the fundamental contributions of Melvin Calvin 
to identifying the reactions concerned this is often referred to as the Calvin 
cycle (see for instance \cite{alberts}). 

In building a model it is necessary to decide which substances are to be 
included. The basic unknowns are the concentrations of these substances. It
is also necessary to decide how the chemical reactions are to be modelled.
In most of this paper diffusion is ignored and only a few remarks are made on
what happens when it is included. In the absence of diffusion the equations 
are of the general form
\begin{equation}
\dot x_i=f_i(x).
\end{equation}
Here $x_i$ are the concentrations, which are functions of time, and 
the dot denotes the time derivative. The solutions of relevance for the 
applications are those for which all $x_i$ are positive. In other words the
point with coordinates $x_i(t)$ is always in the positive orthant $S$ of 
$\R^n$. The mapping $f$ with components $f_i$ represents the interaction 
between the different substances during the reactions. It is of the form 
$Nv(x)$ where $N$ is a matrix called the stoichiometric matrix and the 
components $v_\alpha$ of $v$ are the rates of the different reactions. 
The $v_\alpha$ describe what is called the kinetics. The function $f_i$ 
is of the form $f_i^+-x_if_i^-$ for two non-negative functions $f_i^+$ and 
$f_i^-$ as a result of the form of the dependence of the reaction rates on the 
concentrations of the substances going into the reactions. The cosets of the 
range of the stoichiometric matrix are called stoichiometric 
compatibility classes and are invariant under the flow of the dynamical 
system. In all the systems 
considered in what follows the function $v$ is $C^1$ and it can be shown that 
if the concentrations $x_i$ are positive at some time they remain positive as 
long as the solution exists. This can be proved as in the special case covered 
by Lemma 1 of \cite{rendall11}. Thus $S$ is invariant under the evolution and 
it follows by continuity that its closure $\bar S$ is also invariant. 

A common choice of kinetics is mass action kinetics where if $p$ molecules of 
the substance with concentration 
$x_i$ take part in a reaction the reaction rate has a factor proportional to 
$x_i^p$. This corresponds to the idea that the rate of reaction is 
proportional to the probability of the relevant molecules meeting. For 
instance in the simple reaction $A+2B\to C$ the reaction rate is of the form 
$kx_Ax_B^2$ where $k$ is the reaction constant. For more details on building 
systems of ODE describing reaction networks and mass action kinetics in 
particular see \cite{feinberg80}. Another common choice, particularly in 
the description of biological systems, is Michaelis-Menten kinetics. This is 
adapted to describing reactions which are dependent on a catalyst and the 
reactions in biological systems are usually catalysed by enzymes. If the 
enzymes are explicitly included in the description a type of kinetics is 
obtained which is referred to in \cite{grimbs11} as Michaelis-Menten 
represented in terms of mass action (MM-MA). Michaelis-Menten (MM) kinetics 
is obtained from MM-MA kinetics by a limiting process (quasistationary 
approximation).  

The structure of the paper is as follows. In Section \ref{ma} the dynamics of 
models with mass action kinetics is considered for two different choices of 
the stoichiometric coefficients. In particular it is shown that for certain 
values of the reaction constants there is exactly one positive steady state 
and that it is unstable. This raises the question of the final fate of general
solutions. It turns out that for suitable choices of the reaction constants 
there is an open set of initial data for which all concentrations tend to zero 
at late times and an open set of initial data for which all concentrations tend 
to infinity at late times. The main results are collected in Theorem 1. The 
second statement is somewhat technical to prove for the choice of 
stoichiometric coefficients used in \cite{grimbs11} and the proof is the 
subject of Section \ref{infinity}. Section \ref{mmvma} is concerned with the 
models where the kinetics is Michaelis-Menten represented in terms of mass 
action. It is shown that there are solutions which tend to infinity at late
times for both the choices of stoichiometric coefficients made in 
\cite{grimbs11} and in \cite{zhu09}. For the first case it is proved that there
can exist more than one positive stationary solution in a stoichiometric 
compatibility class and there is some discussion of what happens in the 
second case. The models with Michaelis-Menten kinetics are studied in Section 
\ref{mm} and it is shown that that there are solutions which tend to infinity 
at late times for that model too. It is shown that the stationary solutions 
are essentially the same for the MM-MA and MM models. A model in which the
concentration of ATP is a dynamical variable is discussed in Section \ref{atp}. 
This leads to a system of ODE for which, in contrast to the models discussed 
up to this point, all solutions are bounded in the future. In all this the aim 
is to treat values of the reaction constants which are as general as possible. 
In Section \ref{conclusions} some conclusions are drawn. Appendix A gives an 
introduction to Michaelis-Menten theory. Appendix B collects some technical 
results required for the proofs in the main text.

\section{Mass action kinetics}\label{ma}

This section is mainly concerned with the dynamical system (6) of 
\cite{grimbs11}. There are five variables $x_{\rm RuBP}$, $x_{\rm PGA}$, 
$x_{\rm DPGA}$, $x_{\rm GAP}$ and $x_{\rm Ru5P}$ which are the concentrations of 
the substances abbreviated by the subscripts. They are 
ribulose-1,5-bisphosphate (RuBP), 3-phosphoglycerate (PGA), 
1,3-diphosphoglycerate (DPGA), glyceraldehyde-3-phosphate (GAP) and 
ribulose-5-phosphate (Ru5P). This system has mass action kinetics and is 
called MA in what follows. It is given by
\begin{eqnarray}
&&\frac{dx_{\rm RuBP}}{dt}=k_5x_{\rm Ru5P}-k_1x_{\rm RuBP},\label{ma1}\\
&&\frac{dx_{\rm PGA}}{dt}=2k_1x_{\rm RuBP}-k_2x_{\rm PGA}-k_6x_{\rm PGA},
\label{ma2}\\
&&\frac{dx_{\rm DPGA}}{dt}=k_2x_{\rm PGA}-k_3x_{\rm DPGA},\label{ma3}\\
&&\frac{dx_{\rm GAP}}{dt}=k_3x_{\rm DPGA}-5k_4x_{\rm GAP}^5-k_7x_{\rm GAP},
\label{ma4}\\
&&\frac{dx_{\rm Ru5P}}{dt}=-k_5x_{\rm Ru5P}+3k_4x_{\rm GAP}^5.\label{ma5}
\end{eqnarray}
The $k_i$ are the reaction constants and they are all positive. The alternative 
notation where $(x_{\rm RuBP}, x_{\rm PGA}, x_{\rm DPGA}, x_{\rm GAP},
x_{\rm Ru5P})$ is replaced by $(x_1,x_2,x_3,x_4,x_5)$ is also used.
The state space of interest for the applications is the positive orthant $S$.
Sometimes it is also useful to consider the dynamics on $\bar S$. The origin 
is a stationary solution. The linearization of the system at the origin 
has eigenvalues $(-k_1,-k_2-k_6,-k_3,-k_7,-k_5)$. Thus the origin is a 
hyperbolic sink. Consider a solution which starts at a point of the 
boundary of $S$ other than the origin. Let $N$ be the set of indices $i$
for which the concentration $x_i$ vanishes. Both $N$ and its complement are 
non-empty. Hence there exists $i\notin N$ for which $j\in N$ for 
$j=i+1\ {\rm mod}\ 5$. It follows that $\dot x_j>0$ and so the extension of the 
solution towards the past must lie in the complement of $\bar S$. This implies 
that there exists no solution other than the zero solution which stays in the 
boundary of $S$ for a finite time. This can be used to show that if $x$ is a 
solution which starts in $S$ then its $\omega$-limit set contains no 
point of the boundary of $S$ other than the origin. For suppose that $x^*$ is 
a point of the $\omega$-limit set of $x(t)$ which belongs to the boundary of 
$S$ and is not the origin. Then there is a solution $y$ which passes through 
$x^*$ and lies entirely in the $\omega$-limit set of $x$ and hence in 
$\bar S$. On the other hand it has just been shown that this cannot happen. 
Thus any $\omega$-limit point of a solution starting in $S$ must either 
be the origin or a point of $S$.  

Taking a suitable linear combination of (\ref{ma4}) and (\ref{ma5})
eliminates the nonlinear terms. 
\begin{equation}\label{combi}
\frac{d(3x_{\rm GAP}+5x_{\rm Ru5P})}{dt}=3k_3x_{\rm DPGA}-3k_7x_{\rm GAP}
-5k_5x_{\rm Ru5P}.
\end{equation}
Let $X$ be the maximum of the quantities $x_{\rm RuBP}$, $x_{\rm PGA}$, 
$x_{\rm DPGA}$, $x_{\rm GAP}$ and $3x_{\rm GAP}+5x_{\rm Ru5P}$. Then any solution of 
the system satisfies the integral inequality
\begin{equation}
X(t)\le X(t_0)+C\int_{t_0}^tX(s) ds
\end{equation}
where $C$ is the maximum of $2k_1$, $k_2$, $3k_3$ and $\frac15 k_5$. Thus, by 
Gronwall's inequality, none of the variables can blow up in finite time.
Together with the fact that $\bar S$ is invariant this shows that the 
solution exists globally in the future. Summing up what has been proved so
far gives: 

\noindent
{\bf Proposition 1} A solution of (\ref{ma1})-(\ref{ma5}) with positive 
initial data exists globally to the future, remains positive and has no 
$\omega$-limit points on the boundary of $S$ except possibly the origin. 

It is shown in \cite{grimbs11} that if $k_2>5k_6$ there is a unique stationary 
solution of the system in $S$ and the equilbrium concentrations are calculated 
explicitly in terms of the reaction constants. For this solution
\begin{equation}
x_{\rm RuBP}=k_1^{-1}\left[\frac{k_7^5}{3k_4
\left(\frac{2k_2}{k_2+k_6}-\frac53 \right)^5}\right]^{\frac14}.
\end{equation}
The other equilibrium concentrations can be expressed as
$x_{\rm Ru5P}=\frac{k_1}{k_5}x_{\rm RuBP}$, 
$x_{\rm PGA}=\frac{2k_1}{k_2+k_6}x_{\rm RuBP}$,
$x_{\rm DPGA}=\frac{2k_1k_2}{k_3(k_2+k_6)}x_{\rm RuBP}$,
$x_{\rm GAP}=\left(\frac{k_1}{3k_4}x_{\rm RuBP}\right)^{\frac15}$.
An additional relation which can be derived for the stationary solution is
that $x_{\rm GAP}^4=\frac{k_7(k_2+k_6)}{k_4(k_2-5k_6)}$.
The linearization of the system about the stationary solution has the 
characteristic polynomial
\begin{equation}
(\lambda+k_1)(\lambda+k_2+k_6)(\lambda+k_3)(\lambda+25k_4x_{\rm GAP}^4+k_7)
(\lambda+k_5)-30k_1k_2k_3k_4k_5x_{\rm GAP}^4.
\end{equation}
The constant term is equal to
\begin{equation}
k_1k_3k_5[(k_2+k_6)k_7+5(-k_2+5k_6)k_4x_{\rm GAP}^4]
=-4k_1(k_2+k_6)k_3k_5k_7.
\end{equation}
Because of the signs of the coefficients this polynomial has
exactly one positive root. This means in particular that the stationary 
solution is unstable.  For $k_2\le 5k_6$ there is no stationary solution in
$S$. If $k_2-5k_6$ is allowed to tend to zero while each of 
the reaction constants tends to a non-zero value then the stationary 
point tends to infinity. 

Define a function
\begin{equation}\label{lyapunov1}
L_1=x_{\rm RuBP}+\frac12 x_{\rm PGA}+\frac35 x_{\rm DPGA}+\frac35 x_{\rm GAP}
+x_{\rm Ru5P}.
\end{equation}
Then
\begin{equation}\label{ldot1}
\frac{dL_1}{dt}=-\frac12\left(k_6-\frac15 k_2\right)x_{\rm PGA}
-\frac35 k_7x_{\rm GAP}.
\end{equation}
If $k_2\le 5k_6$ then $L_1$ is a Lyapunov function for the system 
(\ref{ma1})-(\ref{ma5}). This recovers the fact 
that for this parameter range there are no stationary solutions. It fact, when 
combined with Proposition 1, it shows that when that inequality holds all 
solutions converge to the origin as $t\to\infty$ - all concentrations go to 
zero at late times. Thus strong control of the late-time asymptotics of all 
solutions has been obtained in this case.

It remains to consider the case $k_2>5k_6$ where $L_1$ does not seem to give
an interesting conclusion. A useful generalization of $L_1$ is given by
\begin{equation}\label{lyapunov2}
L_2=x_{\rm RuBP}+\frac12 x_{\rm PGA}+\alpha x_{\rm DPGA}+\alpha x_{\rm GAP}
+x_{\rm Ru5P}
\end{equation}
for $\alpha>0$. It satisfies
\begin{equation}\label{ldot2}
\frac{dL_2}{dt}=-\frac12\left[k_6-\left(2\alpha-1\right)k_2\right]
x_{\rm PGA}-[\alpha k_7-(3-5\alpha)k_4x_{\rm GAP}^4]x_{\rm GAP}.
\end{equation}
This means that if $(2\alpha-1)k_2\le k_6$ the function $L_2$ is monotone
decreasing along any solution for which $(3-5\alpha)k_4x_{\rm GAP}^4
<\alpha k_7$. For $\alpha=\frac35$ the function $L_2$ coincides with $L_1$.
Another interesting choice is $\alpha=\frac12$. In that case $L_2$ is 
monotone decreasing on the region defined by the inequality 
$k_4 x_{\rm GAP}^4<k_7$. It follows that if a solution initially satisfies
\begin{equation}
x_{\rm RuBP}+\frac12 (x_{\rm PGA}+x_{\rm DPGA}+x_{\rm GAP})+x_{\rm Ru5P}
< \frac12\left(\frac{k_7}{k_4}\right)^{\frac14}
\end{equation}
then it tends to zero as $t\to\infty$.

It has now been shown that in the case $k_2>5k_6$ there is an open set of 
initial data for which the corresponding solutions tend to the origin as 
$t\to\infty$. The argument just given also provides some information
about the basin of attraction of the origin, which is more than could be
concluded from the fact that the origin is a hyperbolic sink. There is also an 
open set of initial data for which all concentrations $x_i$ tend to infinity 
as $t\to \infty$. The proof of this statement is given in Section 
\ref{infinity}.  

In \cite{zhu09} stoichiometric coefficients are considered which are slightly 
different from those in \cite{grimbs11}. While \cite{zhu09} uses
Michaelis-Menten kinetics it is possible to take mass action kinetics with
the stoichiometric coefficients of \cite{zhu09}. This leads to a system which
is called MAZ in what follows. It differs from the system MA only by the 
facts that the terms $-5k_4x_{\rm GAP}^5$ and $3k_4x_{\rm GAP}^5$ are replaced by 
$-k_4x_{\rm GAP}$ and $\frac35 k_4x_{\rm GAP}$ respectively. In terms of the 
reaction network, the system MA arises from the system MAZ by multiplying the
stoichiometric coefficients in one of the reactions by a constant factor so
that they become integers. For the system MAZ the set $\bar S$ is positively 
invariant. The right hand side of the system is $C^1$ due to the fact that all 
the stoichiometric coefficients on the left hand sides of reactions are 
integers. The arguments used to prove that solutions starting in 
$\bar S$ have no $\omega$-limit points on the boundary of $S$ other than the 
origin for the system MA generalize easily to give the same statements for the 
system MAZ. Since the latter system is linear all solutions exist globally. 
The function $L_1$ of (\ref{lyapunov1}) is a Lyapunov function for the system 
MAZ when $k_2\le 5k_6$ since it also satisfies
the equation (\ref{ldot1}) in this case. It follows that all the statements
about the system MA derived using $L_1$ also hold for the system MAZ. In the
latter case
\begin{equation}
\frac{dL_2}{dt}=-\frac12\left[k_6-\left(2\alpha-1\right)k_2\right]
x_{\rm PGA}-\left[\alpha k_7-\left(\frac 35-\alpha\right)k_4\right]x_{\rm GAP}.
\end{equation}
Thus if $(2\alpha-1)k_2\ge k_6$ and $(3-5\alpha) k_4\ge 5\alpha k_7$ and 
at least one of these two inequalities is strict then $-L_2$ is a 
Lyapunov function. The equations for stationary solutions may be
analysed in this case in a similar way to what was done for the system MA. Four
of the relations obtained are 
$x_{\rm Ru5P}=\frac{k_1}{k_5}x_{\rm RuBP}$, 
$x_{\rm PGA}=\frac{2k_1}{k_2+k_6}x_{\rm RuBP}$,
$x_{\rm DPGA}=\frac{2k_1k_2}{k_3(k_2+k_6)}x_{\rm RuBP}$,
$x_{\rm GAP}=\frac{5k_1}{3k_4}x_{\rm RuBP}$.
Substituting these relations in the remaining equation gives:
\begin{equation}\label{indicator}
5(k_2+k_6)(k_4+k_7)=6k_2k_4.
\end{equation}
Thus there are stationary solutions only when this relation is satisfied
and when it is satisfied there is a whole one-dimensional subspace of them.
This situation is not surprising since the equations are linear in this
case. It is natural to examine the eigenvalues of the matrix on the right
hand side of the equation. The characteristic polynomial is
\begin{equation}
(k_1+\lambda)(k_3+\lambda)(k_5+\lambda)(k_2+k_6+\lambda)(k_4+k_7+\lambda)
-\frac65 k_1k_2k_3k_4k_5.
\end{equation}
As in the case of the system MA all the coefficients in this polynomial are 
positive except possibly for the constant term, which is
\begin{equation}
k_1k_3k_5\left[(k_2+k_6)(k_4+k_7)-\frac65 k_2k_4\right].
\end{equation}
When the expression in brackets is negative there is a positive real eigenvalue 
and there exists at least one solution which tends to infinity as $t\to\infty$
since the origin has a non-trivial linear unstable manifold. If
\begin{equation}
(k_2+k_6)(k_4+k_7)<\frac65 k_2k_4
\end{equation}
then $\alpha$ can be chosen so that $-L_2$ is a Lyapunov function. Hence in 
that case the origin does not belong to the $\omega$-limit point of any 
solution. Since other $\omega$-limit points on the boundary of the positive
orthant have already been ruled out it follows that all solutions tend to 
infinity. The results for the systems MA and MAZ are now collected as
a theorem.

\noindent
{\bf Theorem 1} If $k_2\le 5k_6$ then all solutions of MA and MAZ tend to the 
origin as $t\to\infty$. If $k_2>5k_6$ there is a non-empty open set of initial 
data for MA for which the corresponding solutions tend to the origin as 
$t\to\infty$ and a non-empty open set of initial data for which the 
corresponding solutions tend to infinity as $t\to\infty$. If $k_2>5k_6$ then
there is at least one solution of MAZ which tends to infinity as $t\to\infty$.
If $k_4(k_2-5k_6)>5(k_2+k_6)k_7$ then all solutions of MAZ tend to 
infinity as $t\to\infty$.

\section{Solutions which tend to infinity}\label{infinity}

In this section it is shown that if $k_2>5k_6$ there exist solutions 
of (\ref{ma1})-(\ref{ma5}) for which all $x_i$ tend to infinity as 
$t\to\infty$. In fact there is a non-empty open set of initial data for 
which the corresponding solutions behave in this way. 

\noindent
{\bf Theorem 2} If $k_2>5k_6$ there is a non-empty open set of initial data for 
the system MA for which the corresponding solutions tend to infinity as 
$t\to\infty$ and have the asymptotics
\begin{eqnarray}
&&x_1(t)=Ae^{\alpha t}+\ldots\\
&&x_2(t)=A(V_2/V_1)e^{\alpha t}+\ldots\\
&&x_3(t)=A(V_3/V_1)e^{\alpha t}+\ldots\\
&&x_4(t)=[A(V_3/V_1)\xi_s]^{\frac15}e^{\frac{\alpha t}{5}}+\ldots\\
&&x_5(t)=A(V_5/V_1)e^{\alpha t}+\ldots
\end{eqnarray}
Here $\alpha$, $\xi_s$ and the $V_i$ are fixed positive constants and $A$ is a 
constant depending on the solution.

\noindent{\bf Proof}
For the proof it is useful to introduce the quantity $\xi=\frac{x_4^5}{x_3}$. 
Its evolution equation is given by
\begin{equation}\label{evolxi}
\frac{d\xi}{dt}=5(x_3\xi)^{\frac{4}{5}}\left[k_3-5k_4\xi
+\xi^{\frac15}x_3^{-\frac45}\left(\frac15 k_3-k_7-\frac{k_2x_2}{5x_3}\right)\right].
\end{equation}
Let $\bar x$ be the vector with components $x_i$ for $i=1,2,3,5$. Then four of 
the evolution equations can be rewritten as
\begin{equation}\label{evolnotxi}
\frac{d\bar x}{dt}=M\bar x+R.
\end{equation}
Here the only non-zero component of the vector $R$ is the last one and it is
equal to $3 k_4x_3(\xi-\xi_s)$, $\xi_s=\frac{k_3}{5k_4}$ and
\begin{equation}\label{evolbarx}
M=\left[
{\begin{array}{cccc}
-k_1 & 0 & 0 & k_5 \\ 2k_1 & -(k_2+k_6) & 0 & 0 \\ 
0 & k_2 & -k_3 & 0 \\ 0 & 0 & \frac35 k_3 & -k_5
\end{array}}
\right].
\end{equation}
Equations (\ref{evolxi}) and (\ref{evolnotxi}) are equivalent to the original
system. The solutions to be constructed are obtained as fixed points of a
mapping depending on parameters. To define this mapping some
manipulations of the basic equations (\ref{evolxi}) and (\ref{evolnotxi}) are 
necessary. The equation (\ref{evolxi}) can be rewritten as
\begin{equation}
\frac{d\xi}{dt}=25k_4(x_3\xi)^{\frac45}(\xi_s-\xi)+5\xi\left(
\frac15 k_3-k_7-\frac{k_2x_2}{5x_3}\right).
\end{equation}
The first term on the right hand side can be split into a leading term and a 
remainder with the result that the whole right hand side can be written as
\begin{equation}\label{split}
25k_4(x_3\xi_s)^{\frac45}(\xi_s-\xi)+F(\bar x,\xi)
\end{equation}
where
\begin{equation}\label{Fdef}
F(\bar x,\xi)=25k_4(\bar x_3)^{\frac45}[\xi^{\frac45}-\xi_s^{\frac45}](\xi_s-\xi)
+5\xi\left(\frac15 k_3-k_7-\frac{k_2\bar x_2}{5\bar x_3}\right).
\end{equation}
The equation (\ref{evolnotxi}) can be solved by variation of constants to give
\begin{equation}\label{vcbarx}
\bar x(t)=\exp (Mt)\bar x(0)+\int_0^t\exp(M(t-s))R[\bar x,\xi](s)ds.
\end{equation} 
The function $R[\bar x,\xi](s)$ depends on the functions $\bar x(s)$ and 
$\xi(s)$. 

The matrix $M$ is of the type considered in Appendix 2. Its determinant is
$k_1k_3k_5\left(k_6-\frac15 k_2\right)$. Thus when $k_2>5k_6$ it has exactly 
one positive eigenvalue $\alpha$ and all its other eigenvalues have real 
parts less than $\alpha$. Moreover, there is an eigenvector $V$ with 
eigenvalue $\alpha$ all of whose components are positive. Thus 
\begin{equation}
\exp(Mt)\bar x(0)=\mu_1e^{\alpha t}V+\psi(t;\mu_2,\mu_3,\mu_4)
\end{equation}  
where the components of the function $\psi$ are of the form
\begin{equation}
\psi_i(t;\mu_2,\mu_3,\mu_4)=\sum_{j=2}^4\mu_jW_{ij}
e^{\beta_j t}\sum_{k\ge 0}\nu_{jk}t^k.
\end{equation}
Here the $\beta_j$ are the eigenvalues of $M$ other than $\alpha$, the 
$W_{ij}$ are the components of the eigenvectors other than $V$ 
and the $\nu_{jk}$ are constants which are only non-zero for $k>0$ if the 
eigenvalue $\beta_j$ has multiplicity greater than one. The function $\psi$
depends linearly on the parameters $\mu_2$, $\mu_3$ and $\mu_4$ and satisfies 
$\psi(t;\mu_2,\mu_3,\mu_4)=O(e^{(\alpha-\epsilon)t})$ as $t\to\infty$ for some 
$\epsilon>0$.

From the formula for $\bar x$ we get
\begin{equation}\label{evolx3}
x_3(t)=\mu_1 e^{\alpha t}V_3+e_3\cdot\psi(t;\mu_2,\mu_3,\mu_4)
+\int_0^te_3\cdot\exp(M(t-s))R[\bar x,\xi](s)ds.
\end{equation} 
This can be used to split the quantity $x_3$ in the first term of 
(\ref{split}) into a part containing the leading term in (\ref{evolx3}) and
a remainder term. The result is
\begin{equation}\label{evolxi2}
\frac{d\xi}{dt}=25k_4(\mu_1V_3\xi_se^{\alpha t})^{\frac45}(\xi_s-\xi)+
G[\bar x,\xi,t]
\end{equation}
where
\begin{equation}\label{Gdef}
G[\bar x,\xi,t]=25k_4\xi_s^{\frac45}\left[\bar x_3^{\frac45}
-(\mu_1V_3e^{\alpha t})^{\frac45}\right](\xi_s-\xi)+F(\bar x,\xi).
\end{equation}
The idea is that $G$ contains all the contributions to (\ref{evolxi}) which 
can be considered small.

Equation (\ref{evolxi2}) can also be treated by variation of constants. Let
\begin{equation}
\Phi(t)=\exp\left(-\frac{125k_4}{4\alpha}[(\mu_1V_3\xi_s
e^{\alpha t})^{\frac45}-1]
\right).
\end{equation}
Then
\begin{equation}\label{vcxi}
\xi(t)=\xi_s+\eta_0\Phi(t)+\int_0^t\frac{\Phi (t)}{\Phi(s)}G[\bar x,\xi,s]ds
\end{equation}
with an arbitrary constant $\eta_0$. Finally (\ref{vcbarx}) can be rewritten
as
\begin{equation}\label{vcbarx2}
\bar x(t)=\mu_1e^{\alpha t}V+\psi(t;\mu_2,\mu_3,\mu_4)
+\int_0^t\exp(M(t-s))R[\bar x,\xi](s)ds.
\end{equation}
The integral equations (\ref{vcxi}) and (\ref{vcbarx2}) are those which are
used for the fixed point argument. 

Let $X$ be the set of continuous functions $(\bar x(t),\xi(t))$ defined on the 
interval $[0,\infty)$ which satisfy the inequalities
\begin{equation}\label{basicineq}
\frac{|\bar x(t)-\mu_1 e^{\alpha t}V|}{e^{\alpha t}}\le\delta\mu_1,\ \ \
|\xi(t)-\xi_s|e^{\frac45 \alpha t}\le K
\end{equation} 
for positive constants $\mu_1$, $\delta$ and $K$ which are restricted by some
additional conditions in the fixed point argument. Denote the right hand sides 
of equations (\ref{vcxi}) 
and (\ref{vcbarx2}) by ${\cal T}_1(\xi,\bar x)$ and ${\cal T}_2(\xi,\bar x)$, 
respectively and let ${\cal T}=({\cal T}_1,{\cal T}_2)$. Then $\xi$ and 
$\bar x$ solve (\ref{vcxi}) and (\ref{vcbarx2}) if and only if $(\xi,\bar x)$ 
is a fixed point of ${\cal T}$. In order to ensure that the mapping 
${\cal T}$ is well-defined on $X$ it suffices to assume that 
$\delta\le\frac12\min_i V_i$ and that $K\le\frac12\xi_s$ since these 
inequalities imply the positivity of $\bar x$ and $\xi$. The aim is to show 
that for a suitable choice of the constants $\mu_1$, $\mu_2$, $\mu_3$, 
$\mu_4$, $\eta_0$, $\delta$ and $K$ this rule defines a mapping $\cal T$ from 
$X$ to itself. 

For convenience let ${\cal T}(\xi,\bar x)=(\zeta,y)$. The quantities $\zeta$ 
and $y$ should now be estimated under the assumptions (\ref{basicineq}).
The quantity $\zeta-\xi_s$ is estimated first. It can be written as a sum
$\sum_{i=1}^5 Q_i$ where the individual terms are defined as follows. $Q_1$
denotes the second term on the right hand side of (\ref{vcxi}). $Q_2$ denotes 
the contribution to
the right hand side of (\ref{vcxi}) coming from the first term on the right 
hand side of (\ref{Gdef}). The contribution to the right hand side of 
(\ref{vcxi}) coming from the second term on the right hand side of (\ref{Gdef})
is $Q_3+Q_4+Q_5$. The three summands in this last expression come from three
summands in $F$. The expression $Q_3$ is the contribution coming from the first
term in (\ref{Fdef}). $Q_4$ is the contribution from the first two terms in the
bracket on the right hand side of (\ref{Fdef}) and $Q_5$ is the contribution 
from the third term in that bracket. As a first step towards estimating 
$\zeta-\xi_s$ consider the following identity which holds for any positive 
constants $A$ and $\gamma$.
\begin{equation}\label{identity}
\frac{d}{dt}(\gamma^{-1}A^{-1}e^{-\gamma t}\exp(Ae^{\gamma t}))
=\exp(Ae^{\gamma t})(1-A^{-1}e^{-\gamma t}).
\end{equation} 
If $A\ge 2$ then the second factor on the right hand side of this equation is 
no smaller than one half for any $t\ge 0$. Hence integrating this relation
from $0$ to $t$ gives
\begin{equation}\label{expexp}
\int_0^t\exp(Ae^{\gamma s})ds\le 2\gamma^{-1}A^{-1}e^{-\gamma t}\exp(Ae^{\gamma t}).
\end{equation}
Thus 
\begin{eqnarray}\label{xidiffbound}
&&|\zeta-\xi_s|\le\Phi(t)\left[|\eta_0|+\|G\|_{L^\infty}\int_{0}^t
\frac{1}{\Phi(s)}ds\right]\nonumber\\
&&\le\frac{2}{25k_4}(\mu_1V_3\xi_s)^{-\frac45}\|G\|_{L^\infty}e^{-\frac{4\alpha}{5}t}
+|\eta_0|\Phi(t).
\end{eqnarray} 
The integral has been estimated using (\ref{expexp}) with
$A=\left(\frac{125k_4}{4\alpha}\right)(\mu_1V_3\xi_s)^{\frac45}$ and 
$\gamma=\frac{4\alpha}{5}$. Assuming $\mu_1$ sufficiently large ensures that 
the lower bound on $A$ required for (\ref{expexp}) to hold is satisfied.
Next $G$ is estimated.
\begin{equation}\label{Gbound}
\|G\|_{L^\infty}\le 20 k_4\xi_s^{\frac45}K(\mu_1V_3)^{\frac45}V_3^{-1}
\sup\{1,(V_3-\delta)^{-\frac25}\}\delta+\|F\|_{L^\infty}.
\end{equation}
Choosing $\delta$ small enough allows $Q_2$ to be bounded by 
$\frac{K}{5}e^{-\frac45\alpha t}$. The first term in the expression for $F$ can 
be bounded by
\begin{equation}
Ck_4\mu_1^{\frac45}e^{\frac45\alpha t}(V_3+\delta)^{\frac45}(\xi-\xi_s)^2
\le Ck_4\mu_1^{\frac45}K^2(V_3+\delta)^{\frac45}e^{-\frac{4\alpha}{5}t}
\end{equation}
for a numerical constant $C$. Choosing $K$ sufficiently small allows $Q_3$ 
to be bounded by $\frac{K}{5}e^{-\frac45\alpha t}$. In the second term in $F$ 
the first two contributions can be bounded by $2\xi_s(k_3+5k_7)$. Choose 
$\mu_1$ large enough so that 
\begin{equation}
\frac{4}{5k_4}(\mu_1V_3\xi_s)^{-\frac45}\xi_s(k_3+5k_7)\le K.
\end{equation} 
Then it follows that $Q_4$ is no greater than $\frac{K}{5}e^{-\frac45\alpha t}$. 
The third contribution to $F$ can be bounded by 
$2\xi_sk_2 \left|\frac{x_2}{x_3}\right|$. Now
\begin{equation}
\left|\frac{x_2}{x_3}\right|\le\frac{V_2+\delta e^{-2\alpha t}}
{V_3-\delta e^{-2\alpha t}}.
\end{equation}
To get a lower bound for the denominator in this expression it is assumed
that $\delta\le\frac12 V_3$. Then 
\begin{equation}
\left|\frac{x_2}{x_3}\right|\le\frac{2V_2+V_3}{V_3}.
\end{equation}
If $\mu_1$ is chosen large enough then $Q_5$ is bounded by 
$\frac{K}{5}e^{-\frac45\alpha t}$. To control the 
quantity $Q_1$ assume that $\mu_1$ is so large that 
\begin{equation}
\frac{125k_4}{4\alpha}[(\mu_1 V_3\xi_s)^{\frac45}e^{\frac{4\alpha}{5}t}-1]
\ge\frac{4\alpha}{5}t.
\end{equation} 
for all $t\ge 0$. Then $|\eta_0|\Phi(t)\le |\eta_0|e^{-\frac{4\alpha}{5}t}$. 
Choose $\eta_0$ to be no larger in modulus than $\frac{K}{5}$. Then combining 
the estimates shows that for $K$ and $\delta$ sufficiently small and $\mu_1$
sufficiently large the second defining inequality of the set $X$ is satisfied 
by $\zeta$.

Next $y-\mu_1 e^{\alpha t}V$ is estimated. The function $\psi$ can be bounded by 
an expression of the form $Ce^{(\alpha-\epsilon)t}$ where the constant $C$ depends 
only on the matrix $M$ and $\epsilon$ is a positive constant. Since $\psi$ is 
linear in the parameters $\mu_2$, $\mu_3$ and $\mu_4$ the constant $C$ can be 
made as small as desired by making these parameters small. Thus it can be 
ensured that $e^{-\alpha t}|\psi(t)|\le\frac{\mu_1\delta}{2}$. The integral term 
in (\ref{vcbarx2}) can be estimated by an expression of the form
$C\mu_1 Ke^{\alpha t}$. This can be made as small as desired compared to 
$\delta\mu_1e^{\alpha t}$ by choosing $K$ small. Thus the first defining 
inequality of the set $X$ is satisfied by $y$.

It follows that for a suitable choice of the parameters ${\cal T}$ maps $X$
into itself. Let $\bar x^0(t)=\mu_1e^{\alpha t}V$ and $\xi^0(t)=\xi_s$.
Define a sequence $(\xi^n,\bar x^n)$ recursively by 
$(\xi^{n+1},\bar x^{n+1})={\cal T}((\xi^n,\bar x^n))$. The sequences $\bar x^n$
and $\xi^n$ are uniformly bounded on compact subsets. It then follows from the 
definition of ${\cal T}$ that their time derivatives are uniformly bounded on
compact subsets. By the Arzela-Ascoli theorem there is a subsequence 
$(\xi^{n_r},\bar x^{n_r})$ which converges uniformly to a limit on compact 
subsets. It is possible to pass to the limit in the integral equations 
defining the iteration to see that the limit is a fixed point of ${\cal T}$ 
and hence the desired solution. The solution is uniquely determined by the 
parameters $\mu_i$ and $\eta_0$. The mapping from these parameters to the 
initial data for the solution at $t=0$ is a diffeomorphism onto its image. 
Thus the set of solutions constructed in this way corresponds to an open set 
of initial data. This completes the proof of the theorem.   

\section{Michaelis-Menten via mass action kinetics}\label{mmvma}

Next the system (10) of \cite{grimbs11} is considered. For the convenience
of the reader we reproduce the essential equations here:
\begin{eqnarray}
\dot x_{\rm RuBP}&&=k_{15}x_{\rm Ru5PE_5}-k_1x_{\rm RuBP}x_{\rm E_1}+k_2x_{\rm RuBPE_1},
\label{mmmac1}\\
\dot x_{\rm RuBPE_1}&&=k_1x_{\rm RuBP}x_{\rm E_1}-(k_2+k_3)x_{\rm RuBPE_1},
\label{mmmac2}\\
\dot x_{\rm PGA}&&=2k_3x_{\rm RuBPE_1}-k_4x_{\rm PGA}x_{\rm E_2}
+k_5x_{\rm PGAE_2}\nonumber\\
&&-k_{16}x_{\rm PGA}x_{\rm E_6}+k_{17}x_{\rm PGAE_6},\label{mmmac3}\\
\dot x_{\rm PGAE_2}&&=k_4x_{\rm PGA}x_{\rm E_2}-(k_5+k_6)x_{\rm PGAE_2},\label{mmmac4}\\
\dot x_{\rm DPGA}&&=k_6x_{\rm PGAE_2}-k_7x_{\rm DPGA}x_{\rm E_3}
+k_8x_{\rm DPGAE_3},\label{mmmac5}\\
\dot x_{\rm DPGAE_3}&&=k_7x_{\rm DPGA}x_{\rm E_3}-(k_8+k_9)x_{\rm DPGAE_3},
\label{mmmac6}\\
\dot x_{\rm GAP}&&=k_9x_{\rm DPGAE_3}-5k_{10}x_{\rm GAP}^5x_{\rm E_4}
+5k_{11}x_{\rm GAPE_4}\nonumber\\
&&-k_{19}x_{\rm GAP}x_{\rm E_7}+k_{20}x_{\rm GAPE_7},\label{mmmac7}\\
\dot x_{\rm GAPE_4}&&=k_{10}x_{\rm GAP}^5x_{\rm E_4}-(k_{11}
+k_{12})x_{\rm GAPE_4},\label{mmmac8}\\
\dot x_{\rm Ru5P}&&=-k_{13}x_{\rm Ru5P}x_{\rm E_5}
+k_{14}x_{\rm Ru5PE_5}+3k_{12}x_{\rm GAPE_4},
\label{mmmac9}\\
\dot x_{\rm Ru5PE_5}&&=k_{13}x_{\rm Ru5P}x_{\rm E_5}-(k_{14}+k_{15})x_{\rm Ru5PE_5},
\label{mmmac10}\\
\dot x_{\rm PGAE_6}&&=k_{16}x_{\rm PGA}x_{\rm E_6}-(k_{17}
+k_{18})x_{\rm PGAE_6},\label{mmmac11}\\
\dot x_{\rm GAPE_7}&&=k_{19}x_{\rm GAP}x_{\rm E_7}
-(k_{20}+k_{21})x_{\rm GAPE_7}.\label{mmmac12}
\end{eqnarray}
The equations for the 
concentrations of the free enzymes have been omitted since they can easily be 
reconstructed. As explained below, this could be made into a closed system by 
using the conservation of the total quantity of each enzyme. This has not been 
done so as to prevent the equations becoming even longer. The kinetics is
called Michaelis-Menten represented in terms of mass action in \cite{grimbs11}
and is called the system MM-MA here. The unknowns are of three types. There
are the concentrations of free substrates, which are denoted by the same
variables as in the system MA. There are the concentrations $x_{\rm E_\alpha}$ of 
the seven enzymes corresponding to the seven reactions in the system. Finally,
there are the concentrations of the complexes formed when the enzymes bind
to their substrates. The complex formed when the substrate $A_i$ binds to the
enzyme $E_\alpha$ is denoted in what follows by $A_iE_\alpha$. Since in some
reactions $r(\alpha)>1$ molecules of the substrate bind to the enzyme this
might be denoted instead by $A_i^{r(\alpha)}E_\alpha$. Since, however, the 
exponent $r(\alpha)$ is uniquely determined by $\alpha$ we chose the shorter
notation to prevent certain formulae becoming even more cumbersome than
they already are. It should be warned that the word \lq complex\rq\ is
used in two different ways in the literature on reaction networks. The
first meaning is the one just introduced. The other is a formal linear
combination of the chemical species which is on the left or right hand 
side of a reaction. To distinguish these two concepts in what follows they
will be referred to as 'enzyme-substrate complex' and 'reaction complex'
respectively.
 
The total concentration of each enzyme (free plus bound) is a conserved 
quantity and is denoted by $\rho_\alpha$. It follows as for the system MA that 
the set $\bar S$ is invariant. The question, whether solutions starting in 
$S$ can have $\omega$-limit points on the boundary is a little more 
complicated than for the system MA. Note first that there is a 
five-dimensional set $A_1$ of stationary solutions in $\bar S$ defined by 
setting the concentrations of all enzymes to zero together with those of the 
corresponding complexes. This is the set where all $\rho_\alpha$ are zero.
This set cannot contain any $\omega$-limit point of a solution with positive
initial data, since for a solution of that type the $\rho_\alpha$ are positive.
The conservation of the $\rho_\alpha$ defines invariant affine subspaces of $S$ 
of codimension seven. It is elementary to show that the stoichiometric 
matrix has rank twelve so that these subspaces are the stoichiometric 
compatibility classes. Call one of these subspaces $S_E$. Another set of 
stationary solutions $A_2$, of dimension seven, is defined by setting the 
concentration of all substrates and all enzyme-substrate complexes to zero. Any
subspace $S_E$ intersects this set of stationary solutions in precisely one
point. Consider a solution $x(t)$ which is positive and which has an 
$\omega$-limit point $x^*$ on the boundary of $S$. The solution $y(t)$ passing 
through $x^*$ lies entirely in the boundary of $S$. If the concentration of 
any free enzyme $E_\alpha$ vanishes at $x^*$ then, by the conservation laws, the 
concentration of the corresponding substrate-enzyme complex is non-zero. It 
follows that the time derivative of the concentration of $E_\alpha$ is positive,
a contradiction. Thus it can be concluded that the concentrations of all free 
enzymes are non-zero at $x^*$. Suppose that $x^*\notin A_2$.  If all 
substrates had zero concentration at $x^*$ then at least one enzyme-substrate 
complex would have to be non-zero and the evolution equation for that 
substrate would give a contradiction. Thus at least one substrate must have 
non-zero concentration. Then the evolution equation for a complex of that 
substrate with any enzyme implies that the concentration of that complex must 
be non-zero. Since $x^*$ is on the boundary of $S$ it is not possible that the 
concentrations of all substrates are non-zero. From this point on it is 
possible to argue as in the corresponding proof for the system MA 
to obtain a contradiction. It can be concluded that any $\omega$-limit point 
of a solution starting in $S$ must either be a point of $S$ or belong to 
$A_2$. The conservation laws show that the $\omega$-limit set contains at most 
one point of $A_2$.

The system (\ref{mmmac1})-(\ref{mmmac12}) has the property that all the 
variables 
representing the concentrations of enzymes or enzyme-substrate complexes are 
bounded due to the conservation laws for the quantities $\rho_\alpha$. On the 
other hand, the quantities on the right hand sides of the evolution equations 
for all other variables are all either non-positive or linear in the 
quantities which are not known to be bounded. It follows from this that all 
solutions exist globally in the future. Summing up:

\noindent
{\bf Proposition 2} A solution of the system (\ref{mmmac1})-(\ref{mmmac12})
(the system MM-MA) with positive initial data exists globally to the future, 
remains positive and has no $\omega$-limit points on the boundary of $S$ 
except possibly a single point of the set $A_2$.

\noindent
The conclusions listed for the system MM-MA in this proposition also hold for
the analogous system MM-MAZ defined using the stoichiometric coefficients 
of \cite{zhu09} and can be proved in the same way.

For $i=1,2,3,4,5$ let $\tilde x_i$ be the sum of the concentration of the free 
substrate $i$ and its concentrations within its complexes with enzymes. Note 
that here it is necessary to take into account that in general the complex 
contains several molecules of the substrate. Then $A_2$ is the subset of 
$\bar S$ where all $\bar x_i$ vanish. These quantities satisfy the evolution 
equations
\begin{eqnarray}\label{barx}
&&\frac{d\tilde x_1}{dt}=k_{15}x_{\rm Ru5PE_5}-k_3x_{\rm RuBPE_1},\\
&&\frac{d\tilde x_2}{dt}=2k_3x_{\rm RuBPE_1}-k_6x_{\rm PGAE_2}-k_{18}x_{\rm PGAE_6},\\
&&\frac{d\tilde x_3}{dt}=k_6x_{\rm PGAE_2}-k_9x_{\rm DPGAE_3},\\
&&\frac{d\tilde x_4}{dt}=k_9x_{\rm DPGAE_3}-5k_{12}x_{\rm GAPE_4}-k_{21}x_{\rm GAPE_7},\\
&&\frac{d\tilde x_5}{dt}=3k_{12}x_{\rm GAPE_4}-k_{15}x_{\rm Ru5PE_5}.
\end{eqnarray}
Let $\tilde L_1$ be the quantity obtained by replacing $x_i$ by
$\tilde x_i$ in the expression for the function $L_1$ introduced for the 
system MA. Then
\begin{equation}\label{lyapunovbar}
\frac{d\tilde L_1}{dt}=-\frac12 \left( k_{18}x_{\rm PGAE_6}
-\frac{1}{5}k_6x_{\rm PGAE_2}
\right)-\frac35 k_{21}x_{\rm GAPE_7}.
\end{equation}
This shows that $\tilde L_1$ is a Lyapunov function on the region where the
quantity in brackets in (\ref{lyapunovbar}) is non-negative. It will now be 
shown that this can be used to prove that certain solutions tend to zero as
$t\to\infty$.

\noindent
{\bf Proposition 3} A solution of the system MM-MA (the system 
(\ref{mmmac1})-(\ref{mmmac12})) with $k_{17}+k_{18}<k_5+k_6$ which satisfies 
the inequalities 
(\ref{dominance}) and $k_4k_6\rho_2< k_{16}k_{18}(\rho_6-2\bar L_1(0))$ 
converges to a point of $A_2$ as $t\to\infty$.

\noindent
{\bf Proof} For a solution satisfying the assumptions of the proposition
the quantity in brackets in (\ref{lyapunovbar}) is initially positive, i.e.  
\begin{equation}\label{dominance}
k_6x_{\rm PGAE_2}(0)<5k_{18}x_{\rm PGAE_6}(0).
\end{equation}
The evolution equations for the concentrations occurring in this
inequality are
\begin{eqnarray}
&&\frac{d x_{\rm PGAE_2}}{dt}=k_4x_{\rm PGA}x_{\rm E_2}-(k_5+k_6)x_{\rm PGAE_2},\\
&&\frac{d x_{\rm PGAE_6}}{dt}=k_{16}x_{\rm PGA}x_{\rm E_6}-(k_{17}+k_{18})x_{\rm PGAE_6}.
\end{eqnarray}
Let $t_*$ be supremum of times for which 
$k_6x_{\rm PGAE_2}(t)<5k_{18}x_{\rm PGAE_6}(t)$ 
holds on the interval $[0,t_*)$. Using the fact that $k_{17}+k_{18}<k_5+k_6$ 
the sum of the contributions of the second terms on the right hand sides of the 
evolution equations for $x_{\rm PGAE_2}$ and $x_{\rm PGAE_6}$ to the evolution
equation for $5k_{18}x_{\rm PGAE_6}-k_6x_{\rm PGAE_2}$ is positive when $t=t_*$. Now 
$k_4x_{\rm PGA}x_{\rm E_2}\le k_4x_{\rm PGA}\rho_2$ and 
\begin{equation}\label{E6ineq}
k_{16}x_{\rm PGA}x_{\rm E_6}= k_{16}x_{\rm PGA}(\rho_6-x_{\rm PGAE_6})
\ge k_{16}x_{\rm PGA}(\rho_6-2\tilde L_1(0)).
\end{equation}
Thus due to the inequality $k_4k_6\rho_2<k_{16}k_{18}(\rho_6-2\tilde L_1(0))$ 
the assumption that $t_*$ is finite leads to a contradiction. In addition it 
can be seen that in this case any $\omega$-limit point of the solution must 
satisfy $x_{\rm PGA}=0$ and hence belong to $A_2$. This gives the conclusion of
the proposition.

Next it will be shown that the system MM-MA has solutions for which the 
concentrations of the substrates tend to infinity at late times. To do 
this it is most economical to do the calculations in the framework of
a class of reaction networks wider than those describing the Calvin 
cycle. Consider a system of chemical reactions as defined by sets of species, 
reaction complexes and reactions. This will be called the basic reaction 
network. It is possible to build a new network by replacing each reaction in 
the basic network by a Michaelis-Menten scheme containing a substrate (the 
species from the basic network), an enzyme and a substrate-enzyme complex. 
Applying mass action kinetics to the extended network gives 
\lq Michaelis-Menten expressed in terms of mass action\rq\ kinetics or, for 
short, MM-MA kinetics. In this way starting from the basic network we get a 
system of ordinary differential equations called the MM-MA 
system. It contains reaction constants for each reaction in the extended 
system as parameters. Call the substrates $A_i$ and the enzymes $E_\alpha$ for 
some indices $i$ and $\alpha$. The complex formed when these bind to each 
other is denoted by $A_i E_\alpha$.

Some restrictions will now be made on the basic set of chemical reactions.

\begin{enumerate}
\item Each reaction complex in the basic network contains only one species
\item The set of substrates and the set of enzymes are disjoint. This rules out 
the MAP kinase cascade \cite{huang96}.
\item Each enzyme catalyses only one reaction. This rules out systems with 
enzyme sharing such as the multiple futile cycle \cite{wang08}.
\end{enumerate}
When there are $n$ species and $r$ reactions in the basic network then the 
number of species in the corresponding MM-MA system
is $n+2r$. There are $n$ substrates, $r$ free enzymes and $r$ 
substrate-enzyme complexes. There are $3r$ reactions. In the motivating
example for this work the basic system is defined by the equations 
(\ref{ma1})-(\ref{ma5}) describing the Calvin cycle. It has five species. 
There are seven reactions and so there are nineteen species in the 
corresponding MM-MA system. When a system is so big it is not very efficient 
to write it out explicitly when analysing it. It can be more useful to treat 
it as an example of a class of systems characterized by some particular 
structural properties. This is the motivation for considering a more general 
class of systems here. Note that the first restriction above rules out the 
more detailed models of the Calvin cycle given in \cite{pettersson88} and 
\cite{poolman01}. It also rules out the homogeneous case of the model with 
diffusion considered in \cite{grimbs11}. It will be seen in Section \ref{atp}
that in fact all solutions of the latter system are bounded.

The main theme of what follows is solutions of an MM-MA system in which the 
concentrations of all substrates tend to infinity as $t\to\infty$. In fact 
they all tend to infinity linearly in time. In the solutions of interest here
the concentration of each free enzyme tends to zero as $t\to\infty$ and 
almost all the enzyme becomes bound to the substrate at late times. A class
of networks are considered which are called autocatalytic. It is shown that
for MM-MA systems arising from networks satisfying this additional property,
which is defined later, there are large classes of solutions of the type
just described. They are referred to here as runaway solutions. 

The MM-MA system can be written as a set of evolution equations for the 
substrates, the substrate-enzyme complexes and the free enzymes. The right
hand sides of the equations of the second and third types for a given choice 
of enzyme differ only by an overall sign. Adding them gives a conservation 
law for the total amount of enzyme $\rho_\alpha=x_{\rm A_i E_\alpha}+x_{\rm E_\alpha}$. 
The conservation laws can be used to eliminate the concentrations of 
the free enzymes from the evolution equations for the substrates and the 
substrate-enzyme complexes. The evolution equations for the free enzymes can 
be discarded. This leads to the system
\begin{eqnarray}
&&\frac{dx_{\rm A_m}}{dt}=-\sum_{\alpha:i(\alpha)=m} r(\alpha)C^+(\alpha)
x_{\rm A_{i(\alpha)}}^{r(\alpha)}(\rho_\alpha-x_{\rm A_{i(\alpha)}E_\alpha})\nonumber\\
&&+\sum_{\alpha:i(\alpha)=m}r(\alpha)C^-(\alpha)x_{\rm A_{i(\alpha)}E_\alpha}
+\sum_{\alpha:f(\alpha)=m}s(\alpha)\Gamma(\alpha)x_{\rm A_{i(\alpha)}E_\alpha}
\label{mmma1},\\
&&\frac{dx_{\rm A_{i(\alpha)}E_\alpha}}{dt}=C^+(\alpha)x_{\rm A_{i(\alpha)}}^{r(\alpha)}
(\rho_\alpha-x_{\rm A_{i(\alpha)}E_\alpha})-(C^-(\alpha)+\Gamma (\alpha))
x_{\rm A_{i(\alpha)}E_\alpha}.
\label{mmma2}
\end{eqnarray}
Here $C^+(\alpha)$, $C^-(\alpha)$ and $\Gamma (\alpha)$ are the reaction 
constants for the reactions involving the enzyme $E_\alpha$. The numbers 
$r(\alpha)$ and $s(\alpha)$ are the stoichiometric coefficients of the 
reaction catalysed by $E_\alpha$, referred to for brevity as the reaction 
$\alpha$. The number of molecules of substrate entering the reaction is 
$r(\alpha)$ and the number of molecules of product which result is denoted 
by $s(\alpha)$. In fact we allow $r(\alpha)$ and $s(\alpha)$ to be any real
numbers satisfying the condition $r(\alpha)\ge 1$. This inequality ensures
that the coefficients in the system of ODE are $C^1$. $i(\alpha)$ is the 
index labelling the substrate entering the reaction $\alpha$ and $f(\alpha)$ 
is the index labelling the product of 
that reaction. The full MM-MA system consists of (\ref{mmma1}), (\ref{mmma2})
and evolution equations for the concentrations of the free enzymes. 
Equations (\ref{mmma1}) and (\ref{mmma2}) are equivalent to the full MM-MA
system in the following sense. If a solution of the full MM-MA system is given 
then the conserved quantities $\rho_\alpha$ can be computed. Then the 
concentrations of the substrates and the substrate-enzyme complexes satisfy
(\ref{mmma1}) and (\ref{mmma2}) with those values of the $\rho_\alpha$. 
Conversely, suppose that a solution of (\ref{mmma1}) and (\ref{mmma2}) is
given with certain values of the $\rho_\alpha$ and that the concentration
of $x_{\rm A_{i(\alpha)}E_\alpha}$ is always less than $\rho_\alpha$. Then defining the 
concentrations of the free enzymes by 
$x_{\rm E_\alpha}=\rho_\alpha-x_{\rm A_{i(\alpha)}E_\alpha}$ gives a solution of the full 
MM-MA system. The following linear combination of equations (\ref{mmma1}) and
(\ref{mmma2}) will be useful later.
\begin{eqnarray}\label{mmmacombined}
&&\frac{d}{dt}\left(x_{\rm A_m}+\sum_{\alpha:i(\alpha)=m}
r(\alpha)x_{\rm A_{i(\alpha)}E_\alpha}\right)
\nonumber\\
&&=\sum_{\alpha:i(\alpha)=m}r(\alpha)\Gamma(\alpha)x_{\rm A_{i(\alpha)}E_\alpha}
-\sum_{\alpha:f(\alpha)=m}s(\alpha)\Gamma(\alpha)x_{\rm A_{i(\alpha)}E_\alpha}.
\end{eqnarray}

In order to investigate when the MM-MA system admits runaway solutions a
first step is to look for consistent leading order asymptotics. This is 
done using the following ansatz.
\begin{eqnarray}
&&x_{\rm A_m}=\theta_m t+\ldots,\\
&&x_{\rm E_\alpha}=\eta_\alpha t^{-r(\alpha)}+\ldots.
\end{eqnarray} 
For consistency 
$x_{\rm A_{i(\alpha)}E_\alpha}=\rho_\alpha-\eta_\alpha t^{-r(\alpha)}+\ldots$.
These relations and their formal time derivatives are now inserted into
the evolution equations. Comparing coefficients results in the equations
\begin{eqnarray}
&&\theta_m=-\sum_{\alpha:i(\alpha)=m} r(\alpha)C^+(\alpha)
\theta_{i(\alpha)}^{r(\alpha)}\eta_\alpha\nonumber\\
&&+\sum_{\alpha:i(\alpha)=m}r(\alpha)C^-(\alpha)\rho_\alpha
+\sum_{\alpha:f(\alpha)=m}\Gamma(\alpha)s(\alpha)\rho_\alpha ,\label{compare1}\\
&&0=C^+(\alpha)\theta_{i(\alpha)}^{r(\alpha)}\eta_\alpha
-(C^-(\alpha)+\Gamma (\alpha))\rho_\alpha.\label{compare2}
\end{eqnarray}
Substituting the second equation into the first (or comparing
coefficients in (\ref{mmmacombined})) gives
\begin{equation}\label{compare3}
\theta_m=-\sum_{\alpha:i(\alpha)=m} r(\alpha)\Gamma (\alpha)\rho_\alpha
+\sum_{\alpha:f(\alpha)=m} s(\alpha)\Gamma(\alpha)\rho_\alpha.
\end{equation} 
Since the $\theta_m$ are positive this implies a linear system of 
inequalities for the quantities $\rho_\alpha$. If these inequalities admit
non-trivial solutions then the network is said to be autocatalytic.
For a general network it is not easy to determine whether it is 
autocatalytic. The network of \cite{grimbs11} modelling the Calvin
cycle is easily shown to be autocatalytic. The network obtained by replacing 
the stoichiometric coefficients used in \cite{grimbs11} by those used in 
\cite{zhu09} can be checked to be autocatalytic by an almost identical 
computation. When a network is autocatalytic and the $\rho_\alpha$ satisfy 
suitable inequalities then the constants $\theta_m$ are determined by equation 
(\ref{compare3}) and the constants $\eta_\alpha$ are determined by equation 
(\ref{compare2}).

In order to prove the existence of runaway solutions for autocatalytic MM-MA
systems it is convenient to introduce new variables adapted to the 
expected asymptotics. Define
\begin{eqnarray}
&&x_{\rm A_m}(t)=Z_m(t)(t+R),\\
&&x_{\rm E_\alpha}=\zeta_\alpha (t)(t+R)^{-r(\alpha)}.
\end{eqnarray}
Then the solutions to be constructed should satisfy $Z_m(t)\to\theta_m$ 
and $\zeta_\alpha(t)\to\eta_\alpha$ as $t\to\infty$. The parameter $R\ge 1$ has
been introduced to ensure that the leading terms in the quantities which tend
to zero are already small for $t=0$. Rewriting the evolution equations in 
terms of the new variables leads to the system
\begin{eqnarray}
&&(t+R)\frac{d Z_m}{dt}+Z_m=-\sum_{\alpha:i(\alpha)=m} r(\alpha)C^+(\alpha)
Z_{i(\alpha)}^{r(\alpha)}\zeta_\alpha\nonumber\\
&&+\sum_{\alpha:i(\alpha)=m}r(\alpha)C^-(\alpha)\rho_\alpha
+\sum_{\alpha:f(\alpha)=m}s(\alpha)\Gamma(\alpha)\rho_\alpha\nonumber
-F_m(\zeta_\alpha),\label{transformed1}\\
&&\frac{d \zeta_\alpha}{dt}+C^+(\alpha)(t+R)^{r(\alpha)}Z_{i(\alpha)}^{r(\alpha)}
\zeta_\alpha=r(\alpha)(t+R)^{-1}\zeta_\alpha
\nonumber\\
&&+(t+R)^{r(\alpha)}(C^-(\alpha)+\Gamma(\alpha))\rho_\alpha
-(C^-(\alpha)+\Gamma(\alpha))\zeta_\alpha\label{transformed2}
\end{eqnarray}
where
\begin{eqnarray}
&&F_m(\zeta_\alpha)=\sum_{\alpha:i(\alpha)=m}r(\alpha)C^-(\alpha)\zeta_\alpha
(t+R)^{-r(\alpha)}\nonumber\\
&&+\sum_{\alpha:f(\alpha)=m}s(\alpha)\Gamma (\alpha)\zeta_\alpha
(t+R)^{-r(\alpha)}.\label{defF}
\end{eqnarray}
The main result is

\noindent
{\bf Theorem 3} Let an autocatalytic reaction network be given. Then the 
corresponding MM-MA system can be written in the form 
(\ref{transformed1})-(\ref{defF}) depending on a parameter $R$. Fix the 
values of the reaction constants. Define parameters $\theta_m$ and $\eta_\alpha$ 
by equations (\ref{compare3}) and (\ref{compare2}) and suppose that the 
$\theta_m$ are positive. Then there exist positive constants $K$, $R_0$ and 
$\delta_0$ such that if $R\ge R_0$ and 
\begin{equation}
\sum_m |Z_m(0)-\theta_m|+\sum_\alpha |\zeta_\alpha(0)-\eta_\alpha|\le\delta_0
\end{equation}
then 
\begin{equation}
\sum_m |Z_m(t)-\theta_m|+\sum_\alpha |\zeta_\alpha(t)-\eta_\alpha|\le K\delta_0
\end{equation}
for all $t\ge 0$ and 
\begin{equation}
\lim_{t\to\infty}
\left(\sum_m |Z_m(t)-\theta_m|+\sum_\alpha |\zeta_\alpha(t)-\eta_\alpha|\right)=0.
\end{equation}

\vskip 10pt\noindent
To prove this theorem the first step is to rewrite the evolution equation for 
$\zeta_\alpha$ as an integral equation using variation of constants. 
Define 
\begin{equation}
\Psi_\alpha(s,t)=\exp\left[-\int_s^tC^+(\alpha)(u+R)^{r(\alpha)}
(Z_{i(\alpha)}(u))^{r(\alpha)}du
\right].
\end{equation}
Here the fact that $\Psi_\alpha$ depends on $Z_{i(\alpha)}$ has not been made
explicit in the notation. Then 
\begin{eqnarray}\label{zetaalpha}
\zeta_\alpha(t)=\zeta_\alpha(0)\Psi_\alpha(0,t)
+\int_0^t\Psi_\alpha(s,t)(C^-(\alpha)+\Gamma(\alpha))
\rho_\alpha (s+R)^{r(\alpha)}ds
\nonumber\\
+\int_0^t\Psi_\alpha(s,t)[r(\alpha)(s+R)^{-1}\zeta_\alpha(s)-(C^-(\alpha)
+\Gamma (\alpha))\zeta_\alpha(s)] ds.
\end{eqnarray}
The second term on the right hand side of this equation can be transformed 
using the identity
\begin{equation}
\frac{1}{C^+(\alpha)(Z_{i(\alpha)}(s))^{r(\alpha)}}\frac{d}{ds}(\Psi_\alpha(s,t))
=(s+R)^{r(\alpha)}\Psi_\alpha(s,t)
\end{equation}
and integration by parts. The result is
\begin{eqnarray}\label{zetaalpha2}
\zeta_\alpha(t)&-&\frac{(C^-(\alpha)+\Gamma (\alpha))\rho_\alpha}
{C^+(\alpha)(Z_{i(\alpha)}(t))^{r(\alpha)}}=\left[\zeta_\alpha(0)
-\frac{(C^-(\alpha)+\Gamma (\alpha))\rho_\alpha}
{C^+(\alpha)(Z_{i(\alpha)}(0))^{r(\alpha)}}\right]\Psi_\alpha(0,t)
\nonumber\\
&&+\int_0^t\Psi_\alpha(s,t)\frac{r(\alpha)(C^-(\alpha)
+\Gamma (\alpha))\rho_\alpha}
{C^+(\alpha)(Z_{i(\alpha)}(s))^{r(\alpha)+1}}\frac{dZ_{i(\alpha)} (s)}{ds} ds+\ldots.
\end{eqnarray}
where the last term in (\ref{zetaalpha}) has not been written out again. 

\noindent
{\bf Proof of Theorem 3} In this proof it is assumed that $K$ and $R_0$ are
greater than one. For positive constants $K$ and $\delta$ define
\begin{equation}
t^*=\sup\left\{t>0:\sum_m|Z_m(t)-\theta_m|+\sum_\alpha|\zeta_\alpha(t)-\eta_\alpha|
\le 2K\delta_0\right\}.
\end{equation}
Suppose that $\delta_0\le\frac14 K^{-1}\min\{\theta_m\}$. This implies that
the inequality $Z_m\ge \theta_m/2$ holds on $[0,t^*]$. The time derivatives 
of the quantities $Z_m$ can be bounded by a constant $K$ depending only on the
parameters in the system. To obtain estimates for $\Psi_\alpha$ the 
following auxiliary estimate is useful. For a positive constant $a$
\begin{eqnarray}
&&\int_0^t\exp \left[-a(r(\alpha)+1)\int_s^t (u+R)^{r(\alpha)}du\right]ds
\nonumber\\
&&=\exp [-a(t+R)^{r(\alpha)+1}]\int_0^t\exp [a(s+R)^{r(\alpha)+1}]ds
\nonumber\\
&&=\exp [-a(t+R)^{r(\alpha)+1}]\int_0^t\frac{1}{a(r(\alpha)+1)(s+R)^{r(\alpha)}}
\frac{d}{ds}\left(\exp [a(s+R)^{r(\alpha)+1}]\right) ds
\nonumber\\
&&=\frac{1}{a(r(\alpha)+1)(t+R)^{r(\alpha)}}
-\frac{\exp [aR^{r(\alpha)+1}-a(t+R)^{r(\alpha)+1}]}
{a(r(\alpha)+1)R^{r(\alpha)}}\nonumber\\
&&+\int_0^t\frac{r(\alpha)}{a(r(\alpha)+1)(s+R)^{r(\alpha)}}
\exp [a(s+R)^{r(\alpha)+1}-a(t+R)^{r(\alpha)+1}]ds.
\end{eqnarray}
Choosing $R_0$ large enough ensures that the first factor in the last integral
is smaller than $\frac12$. Thus the integral term can be absorbed into the left
hand side of the inequality. Discarding a term with a good sign gives
\begin{equation}
\int_0^t\exp \left[-a(r(\alpha)+1)\int_s^t (u+R)^{r(\alpha)}du\right]ds
\le \frac{2}{a(r(\alpha)+1)(t+R)^{r(\alpha)}}.
\end{equation}
Making a suitable choice of the constant $a$ leads to the inequality
\begin{equation}\label{powerlaw}
\int_0^t\Psi_\alpha(s,t)ds\le K (t+R)^{-r(\alpha)}.
\end{equation}
Putting this information into (\ref{zetaalpha2}) gives
\begin{equation}\label{zetaestimate}
\left|\zeta_\alpha(t)-\frac{(C^-(\alpha)+\Gamma (\alpha))\rho_\alpha}
{C^+(\alpha)(Z_{i(\alpha)}(t))^{r(\alpha)}}\right|
\le \left[\zeta_\alpha(0)
-\frac{(C^-(\alpha)+\Gamma (\alpha))\rho_\alpha}
{C^+(\alpha)(Z_{i(\alpha)}(0))^{r(\alpha)}}\right]+K (t+R)^{-r(\alpha)}.
\end{equation}
The first term can be bounded using the smallness condition on the initial data
and the second by using (\ref{powerlaw}) and choosing $R_0$ large. It follows
that for $R_0$ sufficiently large and $\delta_0$ sufficiently small
\begin{equation}\label{relative} 
\left|\zeta_\alpha(t)-\frac{(C^-(\alpha)+\Gamma (\alpha))\rho_\alpha}
{C^+(\alpha)(Z_{i(\alpha)}(t))^{r(\alpha)}}\right|
\le\frac{K\delta_0}{2}.
\end{equation}

The evolution equation for $Z_m$ can be rewritten in the form
\begin{eqnarray}\label{Znew}
&&(t+R)\frac{d Z_m}{dt}+(Z_m-\theta_m)=\nonumber\\
&&-\sum_{\alpha:i(\alpha)=m}r(\alpha)C^+(\alpha)Z_{i(\alpha)}^{r(\alpha)}
\left[\zeta_\alpha-\frac{(C^-(\alpha)+\Gamma(\alpha))\rho_\alpha}
{C^+(\alpha)Z_{i(\alpha)}^{r(\alpha)}}\right]\nonumber\\
&&-F_m(\zeta_\alpha).
\end{eqnarray}
The right hand side of (\ref{Znew}) can be bounded by $K\delta_0/2$
after possibly increasing $K$ and $R_0$. Integrating this gives an inequality
of the form
\begin{equation}
(t+R)\sum_m |Z_m-\theta_m|\le (K\delta_0/2)(t+1)
\end{equation}
and hence 
\begin{equation}\label{zbound}
\sum_m |Z_m-\theta_m|\le \frac{K\delta_0}{2}.
\end{equation}
Combining (\ref{zbound}) and (\ref{relative}) gives
\begin{equation}\label{zetabound}
\sum_\alpha |\zeta_\alpha(t)-\eta_\alpha|\le \frac{K\delta_0}{2}.
\end{equation}
It can be concluded that $t^*=\infty$ and the first part of the theorem is
proved. The integrand in the definition of $\Psi_\alpha$ is bounded below by a
positive constant and thus $\Psi_\alpha(0,t)\to 0$ as $t\to\infty$. Combining 
this with (\ref{zetaalpha2}) shows that 
\begin{equation}
\zeta_\alpha(t)-\frac{(C^-(\alpha)+\Gamma (\alpha))\rho_\alpha}
{C^+(\alpha)(Z_{i(\alpha)}(t))^{r(\alpha)}}\to 0 
\end{equation}
as $t\to\infty$. It can then be concluded from (\ref{Znew}) that 
$\frac{d}{dt}((t+R)(Z_m-\theta_m))=o(1)$. Hence
$(t+R)(Z_m-\theta_m)=o(t)$ and $Z_m\to\theta_m$ as $t\to\infty$. Together
with the information we already have this implies that 
$\zeta_\alpha-\eta_\alpha\to 0$ as $t\to\infty$ and this completes the proof of 
the theorem.  

Consider now stationary solutions of MM-MA. Equation (\ref{mmmacombined}) 
implies that the equation obtained by setting $\theta_m=0$ in (\ref{compare3})
holds in the stationary case. This is a linear system for the 
substrate-enzyme complexes. It is a system of $n$ equations for $r$
unknowns. In the case of the system (\ref{mmmac1})-(\ref{mmmac12})
there are five equations for seven unknowns and it is easily seen that the 
solution space is of dimension two. The equations are those obtained by 
setting the time derivatives to zero in (\ref{barx}). Explicitly
\begin{eqnarray}
x_{\rm RuBPE_1}&&=\frac{k_{15}}{k_3}x_{\rm Ru5PE_5},\label{sec1}\\
x_{\rm PGAE_2}&&=\frac{2k_3}{k_6}x_{\rm RuBPE_1}
-\frac{k_{18}}{k_6}x_{\rm PGAE_6},\label{sec2}\\
x_{\rm DPGAE_3}&&=\frac{k_6}{k_9}x_{\rm PGAE_2},\label{sec3}\\
x_{\rm GAP{E_4}}&&=\frac{k_9}{5k_{12}}x_{\rm DPGAE_3}
-\frac{k_{21}}{5k_{12}}x_{\rm GAPE_7},\label{sec4}\\
x_{\rm Ru5PE_5}&&=\frac{3k_{12}}{k_{15}}x_{\rm GAPE_4}.\label{sec5}
\end{eqnarray} 
Suppose now that we prescribe the values of $x_{\rm PGAE_6}$ and $x_{\rm GAPE_7}$.
It is possible to derive the equation 
\begin{equation}
x_{\rm GAPE_4}=\frac{1}{k_{12}}(k_{18}x_{\rm PGAE_6}+k_{21}x_{\rm GAPE_7}).\label{gape4}
\end{equation}
Substituting back into equations (\ref{sec1})-(\ref{sec3}) and (\ref{sec5}) 
gives:
\begin{eqnarray}
x_{\rm RuBPE_1}&&=\frac{3}{k_3}(k_{18}x_{\rm PGAE_6}+k_{21}x_{\rm GAPE_7}),
\label{secxy1}\\
x_{\rm PGAE_2}&&=\frac{1}{k_6}(5k_{18}x_{\rm PGAE_6}
+6k_{21}x_{\rm GAPE_7}),\label{secxy2}\\
x_{\rm DPGAE_3}&&=\frac{1}{k_9}(5k_{18}x_{\rm PGAE_6}+6k_{21}x_{\rm GAPE_7}),
\label{secxy3}\\
x_{\rm Ru5PE_5}&&=\frac{3}{k_{15}}(k_{18}x_{\rm PGAE_6}
+k_{21}x_{\rm GAPE_7}).\label{secxy4}
\end{eqnarray} 
These can then be used to obtain expressions for the concentrations of the 
free enzymes. For the total amount of any one of the enzymes is equal to the 
amount of the free enzyme plus the amount of it bound to its substrate. 
Now these expressions will be used to extract information from the time
evolution equations for the free enzymes. For brevity let
$X=x_{\rm PGAE_6}$ and $Y=x_{\rm GAPE_7}$. Then
\begin{eqnarray}
x_{\rm RuBP}&&=\frac{3(k_2+k_3)(k_{18}X+k_{21}Y)}
{k_1(k_3\rho_1-3k_{18}X-3k_{21}Y)},\label{sxy1}\\
x_{\rm PGA}&&=\frac{(k_5+k_6)(5k_{18}X+6k_{21}Y)}
{k_4(k_6\rho_2-5k_{18}X-6k_{21}Y)},\label{sxy2}\\
x_{\rm DPGA}&&=\frac{(k_8+k_9)(5k_{18}X+6k_{21}Y)}
{k_7(k_9\rho_3-5k_{18}X-6k_{21}Y)},\label{sxy3}\\
x_{\rm GAP}&&=\left[\frac{(k_{11}+k_{12})(k_{18}X+k_{21}Y)}
{k_{10}(k_{12}\rho_4-k_{18}X-k_{21}Y)}\right]^{\frac15},\label{sxy4}\\
x_{\rm Ru5P}&&=\frac{3(k_{14}+k_{15})
(k_{18}X+k_{21}Y)}{k_{13}(k_{15}\rho_5-3k_{18}X-3k_{21}Y)}.\label{sxy5}
\end{eqnarray}
The expressions obtained up to now suffice to determine all unknowns in terms
of $X$, $Y$ and the conserved quantities $\rho_\alpha$. There are, however, 
two further equations which lead to consistency conditions. These are:
\begin{eqnarray}
x_{\rm PGA}&&=\frac{(k_{17}+k_{18})X}{k_{16}(\rho_6-X)},\label{sxy6}\\
x_{\rm GAP}&&=\frac{(k_{20}+k_{21})Y}{k_{19}(\rho_7-Y)}.\label{sxy7}
\end{eqnarray}
Note that equations (\ref{sxy1})-(\ref{sxy7})
only have positive solutions under the restrictions
that $X$ and $Y$ satisfy the inequalities which ensure the positivity of the 
denominators of the right hand sides. Rearranging the equations (\ref{sxy6}) 
and (\ref{sxy7}) gives
\begin{eqnarray}
X&&=\frac{k_{16}\rho_6x_{\rm PGA}}{k_{17}+k_{18}+k_{16}x_{\rm PGA}},\label{X}\\
Y&&=\frac{k_{19}\rho_7x_{\rm GAP}}{k_{20}+k_{21}+k_{19}x_{\rm GAP}}.\label{Y}
\end{eqnarray}
Hence
\begin{equation}
5k_{18}X+6k_{21}Y=\frac{5k_{16}k_{18}\rho_6x_{\rm PGA}}{k_{17}+k_{18}+k_{16}x_{\rm PGA}}
+\frac{6k_{19}k_{21}\rho_7x_{\rm GAP}}{k_{20}+k_{21}+k_{19}x_{\rm GAP}}
\end{equation}
and 
\begin{equation}
k_{18}X+k_{21}Y=\frac{k_{16}k_{18}\rho_6x_{\rm PGA}}{k_{17}+k_{18}+k_{16}x_{\rm PGA}}
+\frac{k_{19}k_{21}\rho_7x_{\rm GAP}}{k_{20}+k_{21}+k_{19}x_{\rm GAP}}.
\end{equation}
Substituting these relations into (\ref{sxy2}) and (\ref{sxy4}) gives
equations of the form:
\begin{eqnarray}
&&x_{\rm PGA}-g_1(x_{\rm PGA},x_{\rm GAP})=0,\label{2dsys1}\\
&&x_{\rm GAP}^5-g_2(x_{\rm PGA},x_{\rm GAP})=0\label{2dsys2}
\end{eqnarray} 
for some rational functions $g_1$ and $g_2$. More explicitly
\begin{eqnarray}
&&g_1(x_{\rm PGA},x_{\rm GAP})=\frac{a_1x_{\rm PGA}+a_2x_{\rm GAP}
+a_3x_{\rm PGA}x_{\rm GAP}}
{b_1+b_2x_{\rm PGA}+b_3x_{\rm GAP}+b_4x_{\rm PGA}x_{\rm GAP}},
\\
&&g_2(x_{\rm PGA},x_{\rm GAP})=\frac{c_1x_{\rm PGA}+c_2x_{\rm GAP}
+c_3x_{\rm PGA}x_{\rm GAP}}
{d_1+d_2x_{\rm PGA}+d_3x_{\rm GAP}+d_4x_{\rm PGA}x_{\rm GAP}}
\end{eqnarray}
for suitable constant coefficients depending on the $k_i$ and the 
$\rho_\alpha$.

\noindent
{\bf Lemma 1} Any positive stationary solution of the system 
(\ref{mmmac1})-(\ref{mmmac12}) defines a positive solution of the system 
(\ref{2dsys1})-(\ref{2dsys2}). Conversely each positive solution of
(\ref{2dsys1})-(\ref{2dsys2}) with given values of $\rho_\alpha$ for which 
the quantities $X$ and $Y$ defined by (\ref{X}) and (\ref{Y}) make the 
denominators in (\ref{sxy1}), (\ref{sxy3}) and (\ref{sxy5}) positive defines 
a positive stationary solution of the system (\ref{mmmac1})-(\ref{mmmac12}).

\noindent
{\bf Proof} The first statement is a direct consequence of the calculations
which have just been done. To prove the converse let 
$(x_{\rm GAP},x_{\rm PGA})$ be a solution of (\ref{2dsys1})-(\ref{2dsys2}) and 
let $X$ and $Y$ be defined by (\ref{X}) and (\ref{Y}). Then (\ref{sxy6}) and 
(\ref{sxy7}) are satisfied. It follows from (\ref{2dsys1})-(\ref{2dsys2})
that (\ref{sxy2}) and (\ref{sxy4}) hold. Next define $x_{\rm RuBP}$, $x_{\rm DPGA}$
and $x_{\rm Ru5P}$ by (\ref{sxy1}), (\ref{sxy3}) and (\ref{sxy5}) respectively. 
Define the quantities $x_{\rm E_i}$ by the conservation laws and the 
quantities $x_{\rm RuBPE_1}$, $x_{\rm PGAE_2}$, $x_{\rm DPGAE_3}$, $x_{\rm GAPE_4}$ and 
$x_{\rm Ru5PE_1}$
by (\ref{gape4}) and (\ref{secxy1})-(\ref{secxy4}). Now all the variables in
the system (\ref{mmmac1})-(\ref{mmmac12}) have been defined and it remains to 
show that they define a stationary solution. Equations (\ref{gape4}) and 
(\ref{secxy1})-(\ref{secxy4}) imply (\ref{sec1})-(\ref{sec5}). At this point
it is useful to think of the system (\ref{mmmac1})-(\ref{mmmac12}) as a special
case of the MM-MA system introduced for a more general class of networks
above. In that framework what has been obtained up to this point in the 
proof is a stationary solution of the equations (\ref{mmma1}), (\ref{mmma2}) 
and (\ref{mmmacombined}). It follows from the discussion above that this 
set of equations is equivalent to the full MM-MA system and this completes
the proof of the lemma.


In \cite{grimbs11} elementary flux modes of this system are investigated.
This is a concept coming from chemical reaction network theory which can 
sometimes be used to investigate the number of stationary solutions of
a dynamical system coming from a network of chemical reactions. To
describe this in more detail it is necessary to introduce the notion of
the deficiency of a reaction network. First note that a directed graph can be 
associated to any network where there is a vertex corresponding to each
reaction complex and an arrow corresponding to each reaction. The 
connected components of this graph are called linkage classes. Let $n$ be 
the number of reaction complexes, $s$ the rank of the stoichiometric matrix 
and $l$ the number of linkage classes. Then the deficiency of the network is
$\delta=n-s-l$. If the deficiency of the network (which is always 
non-negative) is equal to one and some other technical conditions are 
satisfied information about the number of stationary solutions can be obtained 
using what is called the deficiency one algorithm (D1A) \cite{feinberg88}. 
There is a computer implementation of this algorithm which can be applied 
to cases where the network is not too large \cite{ellison00}. An elementary flux
mode defines a subnetwork which is always of deficiency one \cite{conradi07}.
Under suitable technical conditions stationary solutions of the subnetwork
lead to corresponding stationary solutions of the full network and this can
be proved using the implicit function theorem. In \cite{grimbs11} this 
procedure is cited to conclude the existence of two distinct stationary 
solutions of the system MM-MA in a given stoichiometric class. In what follows 
we will not say much more about this approach but the results obtained by 
using it were the starting point of the more direct proof of the existence of
two stationary solutions given here.

In the MM-MA model for the Calvin cycle there are two stoichiometric 
generators and each defines a subnetwork. The subnetwork can
be obtained by setting some of the reaction constants to zero. Here
we procede directly using certain limits for the reaction constants
corresponding to the two modes. Consider the system obtained from the 
system for stationary solutions of the system (\ref{mmmac1})-(\ref{mmmac12}) 
by setting $k_{16}$, $k_{17}$ and $k_{18}$ to zero. Call it LS1. If the limiting 
values of the parameters are approached in such a way that $k_{17}/k_{16}$ and 
$k_{18}/k_{16}$ tend to non-zero limits $q_{17}$ and $q_{18}$ then the system 
varies in a way which is smooth up to the boundary. A similar system LS2 can 
be obtained by letting $k_{19}$, $k_{20}$ and $k_{21}$ tend to zero while 
$k_{20}/k_{19}$ and $k_{21}/k_{19}$ tend to non-zero limits $q_{20}$ and 
$q_{21}$.

\noindent
{\bf Lemma 2} Consider the system LS1 and suppose that 
$k_6\rho_2-6k_{21}\rho_7\ge 0$. If $k_{12}\rho_4-k_{21}\rho_7\ge 0$ then there 
is a unique positive solution. If $k_{12}\rho_4-k_{21}\rho_7<0$ then the number 
of solutions is zero, one or two according to whether 
\begin{equation}
\frac15\left[\frac{4k_{12}\rho_4(k_{20}+k_{21})}
{5k_{19}(k_{12}\rho_4 -k_{21}\rho_7)}\right]^4
k_{10}\rho_4(k_{20}+k_{21})-(k_{11}+k_{12})k_{19}k_{21}\rho_7
\end{equation}
is negative, zero or positive, respectively.
 
\noindent
{\bf Proof} In this case the functions $g_1$ and $g_2$ in (\ref{2dsys1}) and
(\ref{2dsys2}) only depend on $x_{\rm GAP}$. This means that (\ref{2dsys2}) is 
an equation for $x_{\rm GAP}$ alone and for a suitable solution of this 
equation a 
corresponding value of $x_{\rm PGA}$ can be calculated. The explicit form of 
(\ref{2dsys2}) is
\begin{equation}\label{xGAP4}
x_{\rm GAP}^4=\frac{(k_{11}+k_{12})k_{19}k_{21}\rho_7}
{k_{10}k_{12}\rho_4(k_{20}+k_{21})+ k_{10}k_{19}(k_{12}\rho_4 
-k_{21}\rho_7)x_{\rm GAP}}.
\end{equation}
When $k_{12}\rho_4-k_{21}\rho_7\ge 0$ the right hand side of this equation 
is non-increasing and the equation (\ref{xGAP4}) has unique positive 
solution. The explicit form of (\ref{2dsys1}) is
\begin{equation}\label{xPGAs}
x_{\rm PGA}=\frac{6(k_5+k_6)k_{19}k_{21}\rho_7x_{\rm GAP}}
{k_4k_6\rho_2(k_{20}+k_{21})+k_4k_{19}(k_6\rho_2-6k_{21}\rho_7)x_{\rm GAP}}.
\end{equation}
If the right hand side of this is positive it defines an acceptable solution
for $x_{\rm GAP}$ and the first statement of the lemma follows. Consider 
the case $k_{12}\rho_4-k_{21}\rho_7<0$. Rewrite the equation schematically as 
\begin{equation}
x^4=\frac{\alpha}{\beta+\gamma x}.
\end{equation}
where $\alpha>0$, $\beta>0$ and $\gamma<0$. If $p(x)=\gamma x^5+\beta x^4$ 
then the equation to be solved is $p(x)=\alpha$. The function $p$ has a unique 
maximum at $x_*=\frac{4\beta}{-5\gamma}$ and 
$p(x_*)=\frac15\beta\left(\frac{4\beta}{5\gamma}\right)^4$. Comparing this
quantity with $\alpha$ gives the second result of the lemma.

\noindent
{\bf Lemma 3} Consider the system LS2 and suppose that 
$k_{12}\rho_4-k_{18}\rho_6\ge 0$. It has no positive solution if the 
quantities $k_4k_6(k_{17}+k_{18})\rho_2-5(k_5+k_6)k_{16}k_{18}\rho_6$ and
$-k_6\rho_2+5k_{18}\rho_6$ are non-zero with opposite signs and exactly one 
positive solution when they are non-zero and have the same sign. The solution 
is given by
\begin{eqnarray}
&&x_{\rm PGA}=\frac{k_4k_6(k_{17}+k_{18})\rho_2
-5(k_5+k_6)k_{16}k_{18}\rho_6}
{k_4k_{16}(-k_6\rho_2+5k_{18}\rho_6)},\\
&&x_{\rm GAP}^5=\frac{(k_{11}+k_{12})k_{16}k_{18}\rho_6x_{\rm PGA}}
{k_{10}k_{12}\rho_4(k_{17}+k_{18})+k_{10}k_{16}(k_{12}\rho_4-k_{18}\rho_6)x_{\rm PGA}}.
\end{eqnarray} 
The only other case where there exist positive solutions is when  
\begin{equation}
k_4(k_{17}+k_{18})=(k_5+k_6)k_{16}
\end{equation}
and 
\begin{equation}
k_6\rho_2=5k_{18}\rho_6.
\end{equation}
In that case $x_{\rm PGA}$ is arbitrary and there is a continuum of solutions.

\noindent
{\bf Proof} This is a direct calculation.

The parameters which are contained in $g_1$ and $g_2$ are
\begin{equation}
k_4,k_5,k_6,k_{10},k_{11},k_{12},k_{16},k_{17},k_{18},k_{19},k_{20},k_{21},
\rho_2,\rho_4,\rho_6,\rho_7.
\end{equation}
These equations do not depend on $\rho_1$, $\rho_3$ or $\rho_5$. Thus
the conditions required to ensure the positivity of the solutions of 
(\ref{sxy1}), (\ref{sxy3}) and (\ref{sxy5}) can be guaranteed by choosing
$\rho_1$, $\rho_3$ and $\rho_5$ large enough while keeping the other parameters 
fixed. In the two limiting cases considered above these functions simplify.
In each limiting case the functions $g_1$ and $g_2$ depend on only one of the
variables $x_{\rm PGA}$ and $x_{\rm GAP}$. Thus in that case one of the equations 
to
be solved involves only one of the unknowns. If it can be solved then it can
be substituted into the other equation to get the other variable. The 
derivative of the mapping sending the unknowns to the right hand side of the 
two equations is invertible in the two limiting cases. It follows by the 
implicit function theorem that for parameter sets close to those of the 
limiting cases the number of stationary solutions is independent of the values 
of the parameters. It follows in particular that there an open set in the 
space of parameters and conserved quantities for which this construction 
proves the existence of two distinct positive stationary solutions for given
values of the parameters and conserved quantities.

It is possible to define a system MM-MAZ with mass action via Michaelis-Menten 
kinetics starting from the data in \cite{zhu09}. Like the other systems
considered up to now it has the property that $\bar S$ is positively
invariant. In the case of the system MM-MAZ the quantities $\bar x_i$ satisfy 
evolution equations which are the same as those satisfied in the case of the
system MM-MA except that the coefficients $5$ and $3$ are replaced by $1$ and
$\frac35$. Next stationary solutions of MM-MAZ will be considered. The
concentrations of the substrate-enzyme complexes satisfy a system of linear
equations similar to those in the case MM-MA. The relation between these linear
systems can be expressed succinctly by saying that $k_{12}$ is replaced by
$\frac15 k_{12}$. These linear equations can be solved for the concentrations 
of the first five complexes in terms of the other two. The result is similar 
to that for MM-MA with slightly different coefficients. In the equation for
$x_{\rm GAPE_4}$ there is an extra factor of five while the equations for the 
other complexes are as before since they are independent of $k_{12}$.
It is also possible to derive expressions for the concentrations of all
substrates. These are all identical to the corresponding equations in the 
MM-MA case except for the equation for the concentration of GAP. This last
equation is changed in two ways. The first is that $k_{12}$ is replaced by
$\frac15 k_{12}$. The second is that the exponent $\frac15$
is replaced by one. Equations similar to (\ref{2dsys1}) and (\ref{2dsys2}) can 
be derived, with the important difference that the fifth power in 
(\ref{2dsys2}) is replaced by the first power. Thus (\ref{xGAP4}) is
replaced by a linear equation. This linear equation has a unique positive
solution provided a certain sign condition is satisfied and no positive
solution otherwise. Thus for the system MM-MAZ, in contrast to the system
MM-MA, this construction does not lead to a proof of the existence of more
than one stationary solution for any value of the parameters. Setting the 
right hand sides of the equivalents of equations (\ref{sxy2}) and (\ref{sxy4}) 
for the system MM-MA equal to the corresponding quantities coming from the 
right hand sides of equations (\ref{sxy6}) and (\ref{sxy7}) respectively shows 
that the set of $X$ and $Y$ defining stationary solution is the intersection 
of the zero set of two quadratic polynomials in $X$ and $Y$. Hence by 
B\'ezout's theorem \cite{hartshorne77} unless there is a continuum of 
solutions there are at most four.

\section{Michaelis-Menten kinetics}\label{mm}

Starting from the MM-MA system it is possible to obtain a simplified system,
the Michaelis-Menten system (MM system) by passing to a singular limit. This
will now be carried out formally. It will be convenient to describe this in
the context of the more general system introduced in the last section. (The
basic scheme is explained in a simpler case in Appendix A.)
Let $\epsilon$ be a positive real number and define $\tau=\epsilon t$, 
$\tilde x_{\rm A_{i(\alpha)}E_\alpha}=\epsilon^{-1}x_{\rm A_{i(\alpha)}E_\alpha}$ and 
$\tilde x_{\rm E_\alpha}=\epsilon^{-1}x_{\rm E_\alpha}$. Defining  
$\tilde \rho_\alpha=\epsilon^{-1}\rho_\alpha$ allows the transformation of the 
system to be carried out directly on the system (\ref{mmma1})-(\ref{mmma2}).
Equation (\ref{mmma1}) is identical to what it was originally except that the 
original quantities are replaced by the transformed quantities. The factors 
of $\epsilon$ cancel. On the other hand the equation (\ref{mmma2}) picks up
an extra factor of $\epsilon$ on the left hand side. Formally taking the 
limit $\epsilon\to 0$ results in the equation obtained by setting the
time derivative in (\ref{mmma2}) to zero. Solving this equation for 
$\tilde x_{\rm A_{i(\alpha)}E_\alpha}$ gives
\begin{equation}\label{stat}
\tilde x_{\rm A_{i(\alpha)}E_\alpha}=\frac{C^+(\alpha)x_{\rm A_{i(\alpha)}}^{r(\alpha)}
\tilde\rho_\alpha}{C^+(\alpha)x_{\rm A_{i(\alpha)}}^{r(\alpha)}
+C^-(\alpha)+\Gamma(\alpha)}.
\end{equation}
Substituting this back into 
equation (\ref{mmma1}) gives the MM system 
\begin{eqnarray}\label{emm}
&&\frac{dx_{\rm A_m}}{d\tau}=-\sum_{\alpha:i(\alpha)=m}
\frac{C^+(\alpha)\tilde\rho_\alpha
r(\alpha)\Gamma(\alpha)x_{\rm A_{i(\alpha)}}^{r(\alpha)}}
{C^+(\alpha)x_{\rm A_{i(\alpha)}}^{r(\alpha)}+C^-(\alpha)+\Gamma(\alpha)}\nonumber\\
&&+\sum_{\alpha:f(\alpha)=m}\frac{C^+(\alpha)\tilde\rho_\alpha
s(\alpha)\Gamma(\alpha)x_{\rm A_{i(\alpha)}}^{r(\alpha)}}
{C^+(\alpha)x_{\rm A_{i(\alpha)}}^{r(\alpha)}+C^-(\alpha)+\Gamma(\alpha)}.
\end{eqnarray}
Consider the ansatz $x_{\rm A_m}=\tilde\theta_m\tau+...$ for runaway solutions of 
(\ref{emm}). Substituting into the equation and comparing coefficients gives
the equation obtained from (\ref{compare3}) by replacing $\theta_m$ and 
$\rho_{\alpha}$ by $\tilde \theta_m$ and $\tilde\rho_{\alpha}$. Define a new
variable $\tilde Z_m$ by 
\begin{equation}
x_{\rm A_m}(\tau)=\tilde Z_m(\tau)(\tau+R).
\end{equation}
Then equation (\ref{emm}) becomes
\begin{eqnarray}\label{emmZ}
&&(\tau+R)\frac{d\tilde Z_m}{d\tau}+\tilde Z_m=\nonumber\\
&&-\sum_{\alpha:i(\alpha)=m}
\frac{\tilde\rho_\alpha r(\alpha)\Gamma(\alpha)}
{1+(C^+(\alpha))^{-1}\tilde Z_{i(\alpha)}^{-r(\alpha)}
(C^-(\alpha)+\Gamma(\alpha))(\tau+R)^{-r(\alpha)}}
\nonumber\\
&&+\sum_{\alpha:f(\alpha)=m}\frac{\tilde\rho_\alpha s(\alpha)\Gamma(\alpha)}
{1+(C^+(\alpha))^{-1}\tilde Z_{i(\alpha)}^{-r(\alpha)}
(C^-(\alpha)+\Gamma(\alpha))(\tau+R)^{-r(\alpha)}}.
\end{eqnarray}

\noindent
{\bf Theorem 4} Let an autocatalytic reaction network be given. Then the 
corresponding MM system can be written in the form (\ref{emm})
depending on a parameter $R$. Fix the values of the reaction constants. 
The positive parameters $\tilde\theta_m$ are determined by the reaction 
constants. There exist positive constants $K$, $R_0$ and $\delta_0$ 
such that if $R\ge R_0$ and 
\begin{equation}
\sum_m |\tilde Z_m(0)-\tilde\theta_m|\le\delta_0
\end{equation}
then 
\begin{equation}
\sum_m |\tilde Z_m(\tau)-\tilde\theta_m|\le K\delta_0
\end{equation}
for all $\tau\ge 0$ and 
\begin{equation}
\lim_{\tau\to\infty}
\sum_m |\tilde Z_m(\tau)-\tilde\theta_m|=0.
\end{equation}

\noindent
{\bf Proof} The proof is similar to that of Theorem 3 but simpler. Define
\begin{equation}
\tau^*=\sup\left\{\tau>0:\sum_m|\tilde Z_m(\tau)-\tilde\theta_m|\le 
2K\delta_0\right\}.
\end{equation}
Choose $\delta_0$ small enough so that $|\tilde Z_m(\tau)|\ge\tilde\theta/2$ 
on $[0,\tau^*]$. It follows from (\ref{emmZ}) that
\begin{equation}\label{tildeZineq}
\left|(\tau+R)\frac{d\tilde Z_m}{d\tau}+(\tilde Z_m-\tilde\theta_m)\right|
\le K(\tau+R)^{-1}.
\end{equation} 
Integrating this in time and choosing $R_0$ large enough gives an inequality
of the form
\begin{equation}
(\tau+R)(\tilde Z_m-\tilde\theta_m)(\tau)\le K\delta_0 (1+\tau)^\epsilon
\end{equation}   
for any $\epsilon>0$. This allows the bootstrap assumption to be improved if 
$\tau^*$ is finite and it follows that in fact $\tau^*=\infty$. Using 
(\ref{tildeZineq}) again gives the final statement of the theorem. 

Stationary solutions of the MM-MA system give rise to equations similar to 
those defined by the runaway solutions. In that case the analogue of 
(\ref{stat}) without tildes holds. Substituting this into the evolution 
equation for the substrates gives the equation (\ref{emmZ}) without tildes. 
This means that any stationary solution of the MM-MA system defines a 
stationary solution of the MM system. Conversely, any stationary solution of 
the MM system defines a stationary solution of the system 
(\ref{mmma1})-(\ref{mmma2}). It was already shown that any solution of the 
latter system defined a solution of the system MM-MA and if the solution of 
(\ref{mmma1})-(\ref{mmma2}) is stationary the solution of the system MM-MA is 
so too. Thus there is a one to one correspondence between stationary solutions 
of the MM system and stationary solutions of the MM-MA system for fixed values 
of the conserved quantities $\rho_\alpha$. To get the standard form of the 
Michaelis-Menten system as used in \cite{zhu09} the numerators and 
denominators in all the summands on the right hand side should be divided by 
$C^+(\alpha)$. Hence if the MM system is given in isolation the reaction 
constants of the MM-MA system it comes from are not determined uniquely. Only 
the expressions $\Gamma(\alpha)$ and 
$\frac{C^-(\alpha)+\Gamma (\alpha)}{C^+(\alpha)}$ are determined.

Equations for a model with Michaelis-Menten kinetics are written in Appendix 
A of \cite{zhu09}. In fact, as has been remarked in \cite{arnold11}, the
expression for $v_5$ in \cite{zhu09} is not correct since it includes a 
dependence of the reaction on ATP, which does not agree with the reaction it 
is supposed to model, the sixth reaction in Table B2 of \cite{zhu09}. Here we 
consider the correct model obtained from the reaction network given in 
\cite{zhu09} by using Michaelis-Menten kinetics and call it the model MMZ in 
what follows. The discrepancy just mentioned only affects the biological 
interpretation of the constants in the system studied here and not the general 
mathematical form of the equations. The variables used in \cite{zhu09} include 
the concentration of ATP but this is assumed to be constant and so no evolution 
equation is required for it. A system with Michaelis-Menten kinetics is also 
considered (but not written explicitly) in \cite{grimbs11} and is called MM in 
what follows. The only difference between these two systems is that the 
expression $v_4$ in \cite{zhu09} is replaced by an expression of the form 
$\frac{Ax_{\rm GAP}^5}{B+x_{\rm GAP}^5}$ for some positive constants $A$ and $B$. 
For both systems the positive orthant $S$ is invariant. In these 
Michaelis-Menten systems the right hand sides of the evolution equations
are bounded functions of their arguments and so solutions exist globally to
the future. Making use of these facts it can be shown as in the case of 
the systems MA and MAZ that there are no $\omega$-limit points on the boundary
except possibly the origin. To see this it is enough to use the structure of
the first equation in Appendix A of \cite{zhu09}. 

In both cases
\begin{equation}
\frac{dL_1}{dt}=-\frac12(v_5-\frac15v_2)-\frac35 v_6
\end{equation} 
where the $v_j$ denote reaction rates. With the corrected value of $v_5$
we get
\begin{equation}
v_5-\frac15v_2=\frac{V_{5{\rm max}}x_{\rm PGA}}{(x_{\rm PGA}+K_{m5})}
-\frac15\left[\frac{V_{2{\rm max}}x_{\rm PGA}x_{\rm ATP}}
{(x_{\rm PGA}+K_{m21})(x_{\rm ATP}+K_{m22})}\right].
\end{equation}
The positivity of this is equivalent to the inequality
\begin{eqnarray}
&&(x_{\rm PGA}+K_{m5})V_{2{\rm max}}x_{\rm ATP}\nonumber\\
&&\le 5(x_{\rm PGA}+K_{m21})(x_{\rm ATP}+K_{m22})V_{5{\rm max}}.         
\end{eqnarray}
This can be rewritten as
\begin{equation}
x_{\rm PGA}\le\frac{5K_{m21}(x_{\rm ATP}+K_{m22})V_{5{\rm max}}-K_{m5}V_{2{\rm max}}
x_{\rm ATP}}
{V_{2{\rm max}}x_{\rm ATP}-5(x_{\rm ATP}+K_{m22})V_{5{\rm max}}}
\end{equation}
provided the denominator is non-zero. Call the right hand side of this 
equation $K$ and suppose that $K>0$. Then if a solution initially satisfies 
the inequality $\frac53 L_1\le K$ it tends to zero as $t\to\infty$. For the 
values of the parameters given in Appendix B of \cite{zhu09} $K$ is about
$0.14$.

\section{A model including a dynamical description of ATP}\label{atp}

In \cite{grimbs11} a model with diffusion is considered which is given by the 
equations (13) of that reference. They define a system of reaction
diffusion equations. This system is denoted by MAd. Setting the diffusion 
coefficient equal to zero (or restricting to spatially homogeneous solutions)
gives a system of ODE different from (\ref{ma1})-(\ref{ma5}) due to the 
inclusion of the concentration of ATP as a dynamical variable. Call this ODE 
system MAdh. The explicit form of this system is
\begin{eqnarray}
&&\frac{dx_{\rm RuBP}}{dt}=k_5x_{\rm Ru5P}x_{\rm ATP}-k_1x_{\rm RuBP},\label{madh1}\\
&&\frac{dx_{\rm PGA}}{dt}=2k_1x_{\rm RuBP}-k_2x_{\rm PGA}x_{\rm ATP}
-k_6x_{\rm PGA},
\label{madh2}\\
&&\frac{dx_{\rm DPGA}}{dt}=k_2x_{\rm PGA}x_{\rm ATP}-k_3x_{\rm DPGA},\label{madh3}\\
&&\frac{dx_{\rm GAP}}{dt}=k_3x_{\rm DPGA}-5k_4x_{\rm GAP}^5-k_7x_{\rm GAP},
\label{madh4}\\
&&\frac{dx_{\rm Ru5P}}{dt}=-k_5x_{\rm Ru5P}x_{\rm ATP}
+3k_4x_{\rm GAP}^5,\label{madh5}\\
&&\frac{dx_{\rm ATP}}{dt}=-k_2x_{\rm PGA}x_{\rm ATP}-k_5x_{\rm Ru5P}x_{\rm ATP}
+k_8(c-x_{\rm ATP})\label{madh6}
\end{eqnarray}
for a constant $c$. Adding equations (\ref{madh1}) and (\ref{madh6}) gives
\begin{equation}
\frac{d}{dt}(x_{\rm RuBP}+x_{\rm ATP})=-k_1x_{\rm RuBP}-k_2x_{\rm PGA}x_{\rm ATP}
+k_8(c-x_{\rm ATP}).
\end{equation}
Let $m=\min\{k_1,k_8\}$. Then
\begin{equation}
\frac{d}{dt}(x_{\rm RuBP}+x_{\rm ATP})\le -m(x_{\rm RuBP}+x_{\rm ATP})+k_8c.
\end{equation}
It follows that $x_{\rm RuBP}+x_{\rm ATP}$ can be bounded by the maximum of its
initial value and the quantity $\frac{k_8 c}{m}$. Call this $\hat x_{\rm RuBP}$.
The evolution equation for $x_{\rm PGA}$ implies that this quantity is bounded
by the maximum of its initial value and 
$\hat x_{\rm PGA}=\frac{2k_1\hat x_{\rm RuBP}}{k_6}$. Similarly the quantities
$x_{\rm DPGA}$ and $x_{\rm GAP}$ are bounded by the maximum of their initial values
and 
\begin{equation}
\hat x_{\rm DPGA}=\frac{k_2\hat x_{\rm PGA}\hat x_{\rm RuBP}}{k_3},\ \ \ \ \
\hat x_{\rm GAP}=\frac{k_3\hat x_{\rm DPGA}}{k_7},
\end{equation}
respectively. 

Obtaining a bound for $x_{\rm Ru5P}$ is more complicated. 
Integrating the evolution equation
for $x_{\rm DPGA}$ on the interval $[s,t]$ and using the fact that $x_{\rm PGA}$ 
and 
$x_{\rm DPGA}$ are bounded leads to an inequality of the form
\begin{equation}
\int_s^t x_{\rm DPGA}(\xi)d\xi\le C\left(\int_s^t x_{\rm ATP}(\xi)d\xi+1\right)
\end{equation}
for a positive constant $C$. Similarly it follows from the evolution equation 
for $x_{\rm GAP}$ that
\begin{equation}
\int_s^t x_{\rm GAP}(\xi)d\xi\le C\left(\int_s^t x_{\rm DPGA}(\xi)d\xi+1\right).
\end{equation}
Combining these two inequalities gives
\begin{equation}
\int_s^t x_{\rm GAP}(\xi)d\xi\le C\left(\int_s^t x_{\rm ATP}(\xi)d\xi+1\right).
\end{equation}
By variation of constants the evolution equation for $x_{Ru5P}$ implies
\begin{eqnarray}
&&x_{\rm Ru5P}(t)=x_{\rm Ru5P}(0)\exp\left(-k_5\int_0^tx_{\rm ATP}(s)ds\right)
\nonumber\\
&&+3k_4\int_0^t\exp\left(-k_5\int_s^tx_{\rm ATP}(\xi)d\xi\right)(x_{\rm GAP}(s))^5ds.
\end{eqnarray}
for a positive constant $C'$. The first term is bounded and the second can be 
bounded by an expression of the form
\begin{equation}
C\int_0^t\exp\left(-C'\int_s^tx_{\rm GAP}(\xi)d\xi\right)x_{\rm GAP}(s)ds.
\end{equation}
The integrand in the last expression can be written in terms of the
derivative of $\exp\left(-C'\int_s^tx_{\rm GAP}(\xi)d\xi\right)$. Thus this 
expression can be bounded by
\begin{equation}
C\left[1-\exp\left(-C'\int_0^tx_{\rm GAP}(\xi)d\xi\right)\right].
\end{equation}
It follows that $x_{\rm Ru5P}$ is bounded.

Since solutions of the system MAdh are bounded they exist for all future times.
If a solution of MAdh has an $\omega$-limit point on the boundary of $S$ then 
$x_{\rm ATP}$ is positive there. It follows, using the same argument as was 
applied 
to the system MA, that if a solution of MAdh has an $\omega$-limit point on 
the boundary then all concentrations except that of ATP are zero there. The 
evolution equation for $x_{\rm ATP}$ then shows that $x_{\rm ATP}=c$ at that point.
Linearizing about the point $(0,0,0,0,0,c)$ shows that it is a hyperbolic
sink.

Consider the stationary solutions of MAdh, i.e. the homogeneous stationary 
solutions of the MAd model. Combining the first and fifth equations gives 
$x_{\rm GAP}^5=(k_1/3k_4)x_{\rm RuBP}$, just as in the system MA. In fact, as 
shown in \cite{grimbs11}, for a stationary solution all other concentrations 
can be expressed in terms of $x_{\rm RuBP}$. Note first that
\begin{equation}\label{atpeq}
x_{\rm ATP}=c-\frac{8k_1x_{\rm RuBP}}{3k_8}-\frac{k_7}{k_8}
\left(\frac{k_1x_{\rm RuBP}}{3k_4}\right)^{\frac15}.
\end{equation} 
For an admissible solution the right hand of this equation must be positive.
If this is satisfied the concentrations other than $x_{\rm RuBP}$ can be 
computed. The result is that $x_{\rm PGA}=
\frac{2k_1x_{\rm RuBP}}{k_2x_{\rm ATP}+k_6}$,
$x_{\rm DPGA}=\frac{k_7x_{\rm GAP}}{k_3}+\frac{5k_4x_{\rm GAP}^5}{k_3}$,
$x_{\rm Ru5P}=\frac{3k_4x_{\rm GAP}^5}{k_5 x_{\rm ATP}}$. Substituting all these 
relations 
into the evolution equation for PGA gives an equation for $x_{\rm RuBP}$ alone.   
It is of degree ten and hence it is not easy to extract information from it.

An alternative approach is the following. Another equation which can be derived
for stationary solutions of MAdh is
\begin{equation}\label{madhs1}
x_{\rm GAP}=\left[\frac{k_7(k_2x_{\rm ATP}+k_6)}{k_4(k_2x_{\rm ATP}-5k_6)}
\right]^{\frac14}.
\end{equation}
On the other hand, the equation (\ref{atpeq}) can be rewritten as
\begin{equation}\label{madhs2}
x_{\rm ATP}=c-\frac{8k_4}{k_8}x_{\rm GAP}^5-\frac{k_7}{k_8}x_{\rm GAP}.
\end{equation}
Thus we have a set of two equations for the two quantities $x_{\rm GAP}$ and 
$x_{\rm ATP}$
and solving these is equivalent to determining all stationary solutions of 
MAdh. Write the equation (\ref{madhs1}) schematically as 
$x_{\rm GAP}=F_1(x_{\rm ATP})$.
The function $F_1$ is decreasing on the region $x_{\rm ATP}>5k_6/k_2$ where it is 
real. Note that the right hand side of (\ref{madhs2}) is a decreasing function
of $x_{\rm GAP}$ and thus this equation can be inverted to write it 
schematically as $x_{\rm GAP}=F_2(x_{\rm ATP})$ for a decreasing function $F_2$ 
defined on the interval
$[0,c]$ with $F_2(c)=0$. Stationary solutions of MAdh are in one to one 
correspondence with intersections of the graphs of $F_1$ and $F_2$. The
function $F_1$ is strictly convex since $F_1''>0$. On the other hand, the 
function $F_2$ is strictly concave. It follows that the two graphs can 
intersect in at most two points for any given values of the parameters.
If $c<5k_6/k_2$ they do not intersect at all. For fixed values of the reaction
constants if $c$ is sufficiently large the graphs intersect in two points.

For the system MAdh 
\begin{equation}
\frac{dL_1}{dt}=-\frac12\left(k_6-\frac15 k_2x_{\rm ATP}
\right)x_{\rm PGA}-\frac35 k_7x_{\rm GAP}.
\end{equation}
Note that in this system $x_{\rm ATP}$ is bounded by $c$. Hence to make
$L_1$ a Lyapunov function it suffices to require the inequality $ck_2\le 5k_6$.
Thus when this inequality is satisfied all solutions converge to the origin 
as $t\to\infty$. For the system MAdh the function $L_2$ 
with $\alpha=\frac12$ satisfies the same equation as it does for the system 
MA. Hence the same conclusion can be drawn about solutions which converge to 
zero. 

One remark will be made on the behaviour of solutions of systems including a 
diffusion term. In fact we can do this in any space dimension, adding 
diffusion terms in any subset of the equations with any choice of positive 
diffusion constants. Suppose that the domain of the spatial variables is a 
bounded region $\Omega$ with smooth boundary and assume that Neumann boundary 
conditions hold. Let ${\cal L}_1=\int_{\Omega}L_1$. Then
\begin{equation}
\frac{d{\cal L}_1}{dt}=-\int_\Omega\left[\frac12\left(k_6-\frac15 k_2x_{\rm ATP}
\right)x_{\rm PGA}-\frac35 k_7x_{\rm GAP}\right].
\end{equation}
When the inequality $k_6-\frac15 k_2x_{\rm ATP}\ge 0$ holds ${\cal L}_1$ is a 
Lyapunov function. In particular, there are no stationary solutions, 
homogeneous or inhomogeneous, when $ck_2\le 5k_6$. 


\section{Conclusions}\label{conclusions}

An important motivation for the work of \cite{zhu09} and \cite{grimbs11}
on models of the Calvin cycle was to see if photosynthesis can work in 
different steady states. This led to the question of whether the relevant 
mathematical models admit more than one stable positive stationary solution. 
For related work see also \cite{zhu07}, \cite{arnold11}, \cite{jablonsky11} 
and \cite{lei11}.
It was already shown in \cite{grimbs11} that a simple model using mass action
kinetics (the model called MA) never admits even one solution of this type. 
Depending on the values of the reaction constants, either there is no positive 
stationary solution at all or if there is it is unstable. This suggests that 
this is not a very good model. Trying to understand more about what actually 
happens in this model leads to the mathematical question of what the 
solutions departing from a small neighbourhood of the stationary solution 
actually do or, more generally, what the long-time behaviour of solutions is. 
Theorem 1 of this paper provides partial answers to this question. When there 
is no stationary solution the concentrations of all substrates tend to zero. 
When there is a stationary solution some solutions have the property that all 
concentrations tend to zero while others have the property that all 
concentrations tend to infinity (runaway solutions). The latter alternative 
seems to be a further undesirable property of the model. 

The model MM-MA, which has a much larger number of unknowns, does admit more
than one stationary solution for certain values of the reaction constants, as 
was shown in \cite{grimbs11}, and numerical results indicate that one of them 
is stable. In this paper this existence result was reproduced by a more direct
method. The model MM has the same stationary solutions as the model MM-MA
and thus also admits two stationary solutions for certain choices of parameters.
Applying the same method to the related system MMZ with the stoichiometric
coefficients taken from \cite{zhu09} does not reveal the presence of 
multiple stationary solutions and it may be that in that case there are none.
From this point of view the models MM-MA and MM look better that the model MA 
but in fact, as shown in Theorems 3 and 4 of this paper, both the models MM-MA 
and MM exhibit runaway solutions. An 
intuitive explanation for the existence of these solutions is that in all 
these models the concentration of ATP, which is the energy source for the 
reactions, is not modelled explicitly. Instead ATP is implicitly assumed to be 
plentiful and thus present at a constant level. In \cite{grimbs11} another 
model is considered where diffusion is taken into account and the 
concentration of ATP is modelled dynamically. Restricting to spatially 
homogeneous solutions leads to a system of ODE called MAdh. Interestingly, we 
were able to show here that all solutions of MAdh remain bounded, so that 
there are no runaway solutions in that model. Although heuristically plausible 
this is subtle to prove. For suitable values of the parameters this system 
also admits two positive stationary solutions.

There seem to be few general results available on the boundedness of solutions
of systems of ODE arising from chemical reaction networks with mass action
kinetics. One theorem says, using the language of Chemical Reaction Network
Theory, that in a mass action system which is weakly reversible and has
only one linkage class all solutions are bounded \cite{anderson11a}. Neither
of the main hypotheses of that result hold for any of the systems considered
in this paper but perhaps some of the techniques used there might be adapted
to give information about models for the Calvin cycle. For weakly reversible 
systems it might be possible to prove that solutions do not have 
$\omega$-limit points on the boundary of the positive orthant. 
Information about this and relevant references can be found in 
\cite{anderson11b}.  

It is of interest to compare the conditions which allow the conclusions of
the theorems in this paper to be obtained with values of the parameters which 
are biologically reasonable. To do this we start from the biological data 
collected in Appendix B of \cite{zhu09}. In that paper values are given 
relating to Michaelis-Menten kinetics. Assuming that Michaelis-Menten kinetics 
goes over into mass action kinetics when the concentration of the substrate is 
small compared to that of the enzyme it is possible to get values for the 
reaction constant in equations (\ref{ma1})-\ref{ma5}). The results are
\begin{eqnarray}
&&\left(\frac{V_{1{\rm max}}}{K_{m1}},\frac{V_{2{\rm max}}x_{\rm ATP}}{K_{m21}K_{m22}},
\frac{V_{3{\rm max}}}{K_{m3}},\frac{V_{4{\rm max}}}{K_{m4}},
\frac{V_{5{\rm max}}}{K_{m5}},\frac{V_{6{\rm max}}}{K_{m6}},
\frac{V_{13{\rm max}}x_{\rm ATP}}{K_{m131}K_{m132}}\right)\nonumber\\
&&=(3.78, 125, 10.1, 9.63, 4, 0.02, 4.52).
\end{eqnarray}
With these values of the reaction constants the ratio $\frac{k_2}{5k_6}$ which 
plays a key role in Theorem 1 is equal to $1250$. Thus these values are well
within the regime where a positive stationary solution exists.

Is it true that for the system MA with $k_2>5k_6$ every solution either tends
to infinity, the origin or the positive stationary solution? This is not known
but since (\ref{ma1})-(\ref{ma5}) is what is called a monotone cyclic feedback 
system it follows from the main theorem of \cite{malletparet} that bounded 
solutions have $\omega$-limit sets which are no worse that those of a 
two-dimensional system. Using similar ideas it can be shown that almost all
bounded solutions converge to the stationary solution. The system 
(\ref{ma1})-(\ref{ma5}) satisfies $\frac{\partial f_i}{\partial x_j}\ge 0$ for 
$i\ne j$ and is thus cooperative. It is also irreducible in the sense that no 
non-trivial coordinate hyperplane is left invariant by the Jacobian at any 
point. Using this the result on convergence of all bounded solutions 
except for those whose initial conditions belong to a set of measure zero 
follows from a theorem of Hirsch \cite{hirsch85}.

In this paper a variety of different results have been proved about the 
dynamics of solutions of mathematical models for the Calvin cycle. A number
of interesting open questions remain to be investigated. It would be desirable
to have rigorous results on stability of the stationary solutions and the 
(non)-existence of periodic solutions. Inhomogeneous solutions of the system 
with diffusion should be investigated following the initial work in 
\cite{grimbs11}. It would be good to extend the results of this paper to more 
general models of the Calvin cycle such as those of \cite{pettersson88}
and \cite{poolman01}. Finally, the basic motivating question remains: are 
there mathematical models of the Calvin cycle where it can be proved that 
there are at least two homogeneous stable positive stationary solutions?

\appendix
\section{Michaelis-Menten theory}\label{michaelis}

Consider a simple reaction which converts one molecule of the species $S$ (the 
substrate) to one molecule of the substance $P$ (the product). With 
mass action kinetics this leads to the equations $\dot x_S=-kx_S$ and 
$\dot x_P=kx_S$. Suppose now that this reaction is catalysed by an enzyme
$E$. Then there is a reaction with reaction constant $k_1$ in which
the substrate combines with the enzyme to form a complex $SE$. The 
reaction constant for the process of dissociation of $SE$ into $S$ and $E$
will be denoted by $k_{-1}$. Finally there is the reaction in which the 
complex gives rise to the product with reaction constant $k_2$ while setting 
free the enzyme. This gives rise to the system
\begin{eqnarray}
\dot x_S&&=-k_1 x_Sx_E+k_{-1}x_{SE},\\
\dot x_{SE}&&=k_1x_Sx_E-(k_{-1}+k_2)x_{SE},\\
\dot x_{E}&&=-k_1x_Sx_E+(k_{-1}+k_2)x_{SE},\\
\dot x_P&&=k_2x_{SE}.
\end{eqnarray} 
The first three of these equations form a closed system and thus it is 
natural to analyse it first and use the last equation to determine the 
evolution of the concentration of the product afterwards, if desired.
The above system is the MM-MA version of the original simple reaction.
The Michaelis-Menten kinetics will now be derived on a heuristic level.
Note first that the quantity $x_{SE}+x_E$ is conserved. Call it $E_0$.
Substituting the relation $x_E=E_0-x_{SE}$ into the first two evolution 
equations gives a closed system for $x_S$ and $x_{SE}$:
\begin{eqnarray}\label{closed}
\dot x_S&&=-k_1 E_0x_S+(k_{-1}+k_1 x_S)x_{SE},\\
\dot x_{SE}&&=k_1E_0x_S-(k_{-1}+k_1 x_S+k_2)x_{SE}.
\end{eqnarray}
Now introduce $\tau=\epsilon t$, $\tilde x_{SE}=\epsilon^{-1}x_{SE}$ and 
$\tilde E_0=\epsilon^{-1}E_0$ for a constant $\epsilon$. This gives
\begin{eqnarray}
x'_S&&=-k_1 \tilde E_0 x_S+(k_{-1}+k_1 x_S)\tilde x_{SE},\\
\epsilon\tilde x'_{SE}&&=k_1 \tilde E_0 x_S-(k_{-1}+k_1 x_S+k_2)\tilde x_{SE}
\end{eqnarray}
where the primes denote derivative with respect to $\tau$. In the last system
it is possibly to formally pass to the limit $\epsilon\to 0$, corresponding to
a very small amount of enzyme. In the limit the second equation reduces to the 
algebraic equation
\begin{equation}\label{algebraic}
\tilde x_{SE}=\frac{k_1\tilde E_0 x_S}{k_{-1}+k_1 x_S+k_2}.
\end{equation}
Substituting this back into the evolution equation for $x_S$ and gives the 
effective Michaelis-Menten equation
\begin{equation}\label{emma}
x_S'=-\frac{k_1k_2\tilde E_0 x_S}{k_1x_S+k_{-1}+k_2}.
\end{equation} 
It can then be computed that in this set-up $x_P'=-x_S'$.

This type of discussion is quite standard and the reason it is reproduced 
here is to illuminate the relations between the three types of kinetics (MA, 
MM-MA and MM) by an explanation of the simplest example. There is a one to one 
correspondence between stationary solutions of the systems MM-MA and MM, as 
will now be shown. If a stationary solution $(x_S,x_{SE})$ of the system MM-MA 
is given then the equation (\ref{algebraic}) is satisfied. Hence the equation 
(\ref{emma}) holds and a stationary solution of the system MM is obtained. 
Conversely, suppose that a solution $(x_S,\tilde x_{SE})$ is given. Then a 
stationary solution of the system (\ref{closed}) is obtained. Defining 
$x_E=E_0-x_{SE}$ for a positive constant $E_0$ completes it to a stationary 
solution of the system MM-MA.  

\section{A special class of matrices}\label{matrices}

This appendix is concerned with the algebraic properties of some matrices
of a special form which appear in this paper. Let
$A$ be an $n\times n$ matrix with entries $a_{ij}$. Suppose that
$a_{ii}<0$ for each $i$, that $a_{ij}>0$ for $j=i-1\ {\rm mod}\ n$ and 
that $a_{ij}=0$ otherwise. Suppose further that $(-1)^{n+1}\det A>0$. The matrix 
$A+\lambda I$ is positive for a sufficiently large real number $\lambda$, i.e
all its elements are positive. By the Perron-Frobenius theorem \cite{meyer00}
it has a unique eigendirection spanned by a positive vector. Let $p$ be an 
eigenvector of this type with components $p_i$. The corresponding eigenvalue 
is positive. Let it be denoted by $\beta$. Another consequence of the 
Perron-Frobenius theorem is that all other eigenvalues of 
$A+\lambda I$ have modulus smaller than $\beta$. In particular the real part 
of any other eigenvalue is smaller than $\beta$. The vector $p$ is an 
eigenvector of $A$ with eigenvalue $\alpha=\beta-\lambda$ and all other 
eigenvalues of $A$ have real part smaller than $\alpha$. 

If $A$ is a matrix of the above special form then it can be shown that the 
matrix $B=A^{-1}$ is a positive matrix. One way of proving this as follows.
Let $x$ be a vector in $\R^n$ and consider the equation $Ax=y$. Inverting
the matrix is equivalent to solving this equation for $x$. The equation
can be written in components as
\begin{equation}
a_{ii}x_i+a_{i,i-1}x_{i-1}=y_i; 1\le i\le n
\end{equation}
where the indices are to be interpreted modulo $n$. Hence
\begin{equation}
a_{i+1,i}x_i=(-a_{i+1,i+1})x_{i+1}+y_{i+1}.
\end{equation}
Note that the coefficients in this equation are positive. By substituting these
equations into each other successively with $i$ increasing from one to $n$ it 
is possible to obtain an equation of the form:
\begin{equation}
\left(\prod_ia_{i,i+1}\right) x_1=\left(\prod_i(-a_{ii})\right) x_n+\sum c_iy_i 
\end{equation}
where the coefficients $c_i$ are positive. Rearranging this gives
\begin{equation}
(-1)^{n+1}(\det A)x_n=\sum c_iy_i.
\end{equation}
Any other $x_i$ can be determined in an analogous way. The determinant of 
$A$ is equal to $\prod_i a_{ii}+(-1)^{n+1}\prod_i a_{i,i+1}$. Putting these
facts together gives the proof of the desired statement.

It can be concluded from the Perron-Frobenius theorem that $B$ has a unique 
eigendirection spanned by a positive vector. This is also an eigenvector of 
$A$ and so must be proportional to $p$. The corresponding eigenvalue is 
$\alpha^{-1}$ and is positive. Hence $\alpha$ is positive.

\vskip 10pt\noindent
{\it Acknowledgements} One of the authors (ADR) is grateful to Zoran Nikoloski 
for arousing his interest in this subject and for helpful discussions. This
research was partially supported by the Hausdorff Center for Mathematics of
the University of Bonn.

\end{document}